\documentclass[10pt,journal,final]{IEEEtran}
\IEEEoverridecommandlockouts
\newtheorem{theorem}{Theorem}

\usepackage{amsfonts,amssymb}
\usepackage{ifpdf}
\usepackage{cite}
\usepackage{stfloats}
\usepackage{bm}
\usepackage{color,xcolor}
\usepackage{subfigure}
\ifCLASSINFOpdf
   \usepackage[pdftex]{graphicx}
   \graphicspath{{../pdf/}{../jpeg/}}
   \DeclareGraphicsExtensions{.pdf,.jpeg,.png}
\else
   \usepackage[dvips]{graphicx}

   \graphicspath{{../eps/}}

   \DeclareGraphicsExtensions{.eps}
\fi

\usepackage[cmex10]{amsmath}

\usepackage{algorithm}
\usepackage{algorithmic}
\ifCLASSOPTIONcompsoc
\else
  \usepackage[caption=false,font=footnotesize]{subfig}
\fi
\usepackage{cite}
\usepackage{makecell}
\begin{document}
\title{3D UAV Trajectory and Communication Design  for Simultaneous Uplink and Downlink Transmission}

\author{Meng~Hua,~\IEEEmembership{Student Member,~IEEE,}
Luxi~Yang,~\IEEEmembership{Senior Member,~IEEE,}\\
Qingqing~Wu,~\IEEEmembership{Member,~IEEE,}
and~A. Lee Swindlehurst,~\IEEEmembership{Fellow,~IEEE}

\thanks{This paper has been accepted by IEEE Transactions on Communications. Manuscript received January    02, 2020; revised April   27, and accepted June  12, 2020. This work was supported by the National Natural Science Foundation of China under Grants U1936201 and 61971128, Scientific Research Foundation of Graduate School of Southeast University  under Grand  YBPY1859, and the National Key Research and Development Program of China under Grant SQ2019YFB180141-01.
The associate editor coordinating the review of this paper and approving it for publication was Mehdi Bennis. (\emph{Corresponding author: Luxi Yang}.)}
\thanks{M. Hua,  and L. Yang are with the School of Information Science and Engineering, Southeast University, Nanjing 210096, China (e-mail: \{mhua, lxyang\}@seu.edu.cn).}
\thanks{Q. Wu is with the State Key Laboratory of Internet of Things for Smart City and Department of Electrical and Computer Engineering, University of Macau, Macao, China (email: qingqingwu@um.edu.mo).}
\thanks{A. L. Swindlehurst is with the Center for Pervasive Communications and Computing, University of California at Irvine, Irvine, CA 92697 USA (e-mail: swindle@uci.edu).}
}
\maketitle
\vspace{-3em}
\begin{abstract}
In this paper, we  investigate the  unmanned aerial vehicle (UAV)-aided simultaneous uplink and downlink transmission  networks, where one UAV acting as a disseminator  is  connected to  multiple access points (AP), and the other UAV acting as a  base station (BS) collects  data from  numerous sensor nodes (SNs). The goal of this paper is to maximize the  system throughput  by jointly optimizing the 3D UAV trajectory, communication scheduling, and UAV-AP/SN transmit power. We first consider  a special case where  the UAV-BS and UAV-AP trajectories are pre-determined. Although the resulting problem is an integer and non-convex optimization problem, a globally optimal solution is obtained by applying the polyblock outer approximation (POA) method based on  the problem's   hidden monotonic structure. Subsequently, for the general case  considering the  3D UAV trajectory optimization, an efficient iterative algorithm is proposed to alternately optimize the divided sub-problems based on the  successive convex approximation (SCA) technique. Numerical results demonstrate that the proposed design is able to achieve significant system throughput gain over the benchmarks. In addition, the SCA-based method can achieve nearly the  same performance   as the POA-based method  with much lower  computational complexity.
\end{abstract}


\section{Introduction}
With  continuing communication device miniaturization and the increased endurance of  unmanned aerial vehicles (UAVs), new civilian-use markets are emerging for UAVs beyond military applications, including examples such as emergency search, forest fire detection, cargo transport, etc.  Particularly, UAVs are envisioned as a key component of future wireless network technologies that will expand  network coverage and improve system throughput \cite{Mozaffari2019atutorial,zeng2016wireless,azari2018key}. Compared with  terrestrial base stations (BSs) whose the locations are pre-determined and fixed,  UAVs can adaptively control its position  to react as needed to requests for  on-demand services \cite{jiang2012optimization,han2009optimization,zhang2019securing,wu2018common,song2018joint,li2019near,zhou2019uav}.

There are two main paradigms for the integration of UAVs in the  traditional networks, namely UAV-aided wireless networks and cellular-connected UAV networks \cite{zeng2019accessing,jawhar2014framework,dong2014uav,zhan2018energy,zeng2018cellular,zhang2018cellular,Zhangcellular,fotuhi2019survey}. In  the UAV-aided wireless communication scenario,  the  UAV generally  acts as a mobile BS  equipped with a communication transceiver to provide seamless wireless services or to collect the data  from the  ground nodes. UAVs are especially well suited for data collection in sensor networks where the nodes are widely dispersed over a large area \cite{zeng2019accessing,jawhar2014framework,dong2014uav,zhan2018energy}. The sensor nodes (SNs) are typically battery operated, and cannot transmit continuously. Rather than installing dedicated infrastructure, in delay-tolerant applications it is more cost effective to deploy UAVs to visit the SNs and collect the data in a sense-and-carry fashion. In the cellular-connected UAV communication networks, UAVs are regarded as new aerial users that access the cellular network from the sky for  communications \cite{zeng2018cellular,zhang2018cellular,Zhangcellular,fotuhi2019survey}. In  \cite{azari2019cellular}, an in-depth analysis of integration of cellular-connected UAV into the existing wireless  networks is provided from the perspectives of  multiple metrics. Such UAVs can achieve high data rates with low latency due to the high probability of dominant line-of-sight (LoS) propagation paths with its communication targets.


Despite  promising opportunities for UAVs like those  mentioned above, some key challenges remain to be addressed  in order to effectively use them to realize seamless connectivity and  ultra reliable communication in the future. Recently,   UAV deployment and trajectory designs for sensing and communications have received great attention \cite{alzenad20173,al2014optimal,azari2018ultra,mozaffari2016efficient,wang2018joint,dai2019deploy,zeng2017energy,you20193d,sun2019optimal,wu2018Joint,hua2019energy}. The  deployment of a single UAV was investigated in \cite{alzenad20173,al2014optimal,azari2018ultra}, especially \cite{azari2018ultra} derived analytical expressions for the optimal UAV altitude that minimizes the system  outage probability by using stochastic geometry theory. The deployment of multiple UAVs for either  maximizing the  coverage area or system throughput was investigated in \cite{mozaffari2016efficient} and \cite{wang2018joint}, respectively.
The initial 2D  UAV trajectory optimization was  studied  in \cite{zeng2017energy}, where the authors divided the continuous trajectory into multiple discrete segments and solved the discrete problem by convex optimization techniques. Then, 3D trajectory design has been studied in \cite{you20193d} and \cite{sun2019optimal}. The goal of \cite{you20193d}  was to maximize the minimum average data collection rate from all SNs by optimizing the 3D UAV trajectory under  the assumption of Rician fading channels, while the optimal 3D trajectory was obtained by applying  monotonic optimization theory in \cite{sun2019optimal}.  The problem of  multiple UAVs simultaneously serving multiple SNs was first studied in \cite{wu2018Joint} and \cite{hua2019energy}. In \cite{wu2018Joint}, the UAV transmit power and trajectory were  optimized to alleviate the interference received by the SNs and  maximize the minimum achieved rate from all the SNs. In \cite{hua2019energy}, the authors studied  multiple-UAV cooperative secure   transmission problem  by jointly optimizing the UAV trajectory and transmit power.

While the above work has  studied  the typical UAV-aided wireless communication  network either in the uplink transmission or downlink transmission, question of how to integrate the operation of simultaneous uplink and downlink transmission  has not been addressed and remains an open problem. To fill this gap, we study a general heterogeneous network that consists of these two networks.  For the downlink transmission network,   UAV acts as a disseminator, referred to as  UAV-AP,  to disseminate data to the ground access point (AP) (Note that the  AP is also a type of  SNs, we name it as AP to distinguish uplink SNs).  For the UAV-BS based network, the UAV acts as a mobile base station (BS),  referred to as  UAV-BS,  to collect data from the uplink SNs. We aim  to maximize the sum  system throughput, including contributions from both UAV-BS and UAV-AP operations, by jointly optimizing the 3D UAV-BS/UAV-AP trajectory, communication scheduling, and UAV-AP/SN transmit power. We propose an efficient iterative algorithm to address the problem and obtain  a locally optimal solution. In addition, for the special case where  both UAVs' trajectories are pre-determined, we obtain a globally optimal solution by applying  monotonic optimization theory.

\begin{figure}[!t]
\centerline{\includegraphics[width=2.5in]{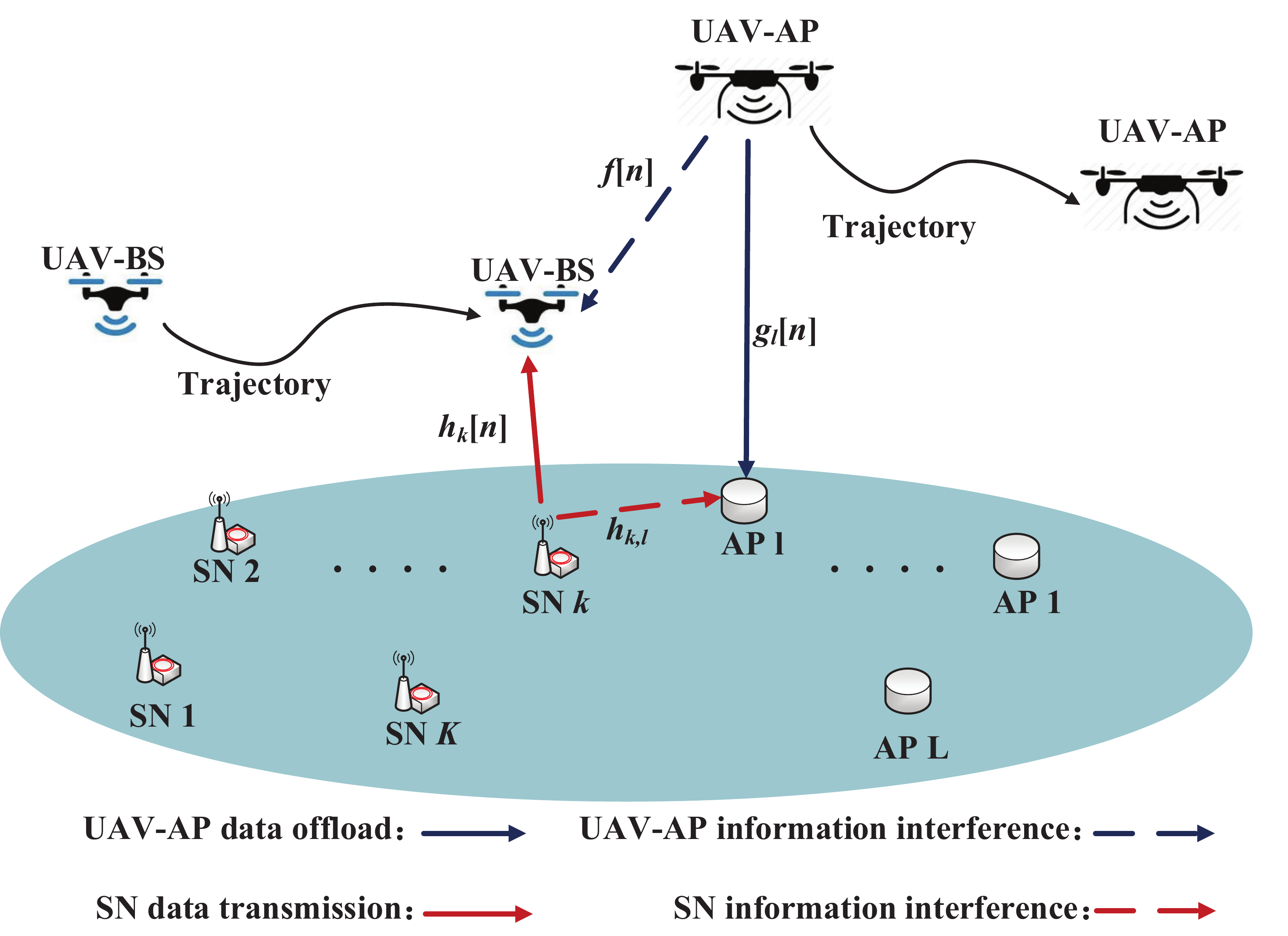}}
\caption{Coexistence of  aerial and cellular-connected UAV networks.} \label{fig1}
\end{figure}

As shown in Fig.~\ref{fig1}, several challenges must be addressed in order to achieve good performance for the  simultaneous uplink and downlink transmission with the help of UAVs.  First, in the UAV-AP based network, namely  downlink transmission, the AP  not only receives the desired signal from the UAV-AP but also suffers from  interference from SNs. Second, in the UAV-BS based network, namely  uplink transmission, the UAV-BS not only collects desired data from the SNs but also encounters interference from the UAV-AP. To enhance system performance, the UAV-BS/UAV-AP trajectories must be carefully designed since the UAV location determines its ability to mitigate interference and increase throughput. Furthermore, transmission power of  UAV-AP  and SN should be jointly optimized to alleviate the whole system interference. Note that this work is different from  \cite{hua2019fulluav}, where a single full-duplex UAV is used to transmit data to the downlink users and receive data from the uplink users simultaneously via a $2\rm D$ trajectory design.  However,  self-interference and multiple access delay issues impede its application for the single full-duplex UAV used in delay-sensitive  tasks.  In this paper, we consider multiple half-duplex  UAVs  to simultaneously serve  downlink users and uplink users, the optimization  of  altitude and transmission power of UAVs resulting in   a heterogeneous networks, which provides additional degrees of freedom for achieving low delay and ultra-reliable communications via  a $3\rm D$ trajectory design. In addition, we propose  a novel method to address the resulting problem, and  a globally optimal solution  is obtained here. It is worth pointing out that work \cite{wu2018Joint} only focuses the case of   multiple UAVs serving multiple users in downlink transmission, whereas the uplink transmission is not considered. To the best of our knowledge, this work is first to study simultaneous uplink and downlink transmission with help of multiple  UAVs.
Our main contributions are  summarized as follows.
\begin{itemize}
\item We  investigate the scenario of simultaneous uplink and downlink transmission with help of multiple UAVs. We  focus on  maximizing the sum of the  UAV-BS and UAV-AP based network throughput subject to  the constraints of UAV mobility  and  SN/UAV-AP transmit power.

\item We first study the case that the UAV-BS and UAV-AP trajectories are pre-determined. We aim at maximizing the  sum system throughput by jointly optimizing the SN/UAV-AP transmit power and communication scheduling. The resulting optimization problem is a non-convex integer optimization problem, whose solution is difficult to obtain. However, by exploiting the hidden monotonic nature of the problem, we find a globally optimal solution  using the polyblock outer approximation  (POA)  method. Note that although  \cite{sun2019optimal} obtains  a globally optimal  solution to solar-powered UAV systems using POA method, it only focuses on  a single UAV in the downlink transmission, we extend it to a more general case with multiple UAVs in the simultaneous uplink and  downlink transmission. In addition, we also propose a suboptimal solution  based on the successive convex approximation (SCA) technique. Our numerical results show that  the SCA-based method can achieve nearly the same system performance  as the POA-based method but  with much lower computational complexity.

\item We then study a more general scenario in which  the UAV-BS and UAV-AP trajectories are optimized. Our goal is to maximize the  sum system throughput by jointly designing  the UAV-BS/UAV-AP trajectories, SN/UAV-AP transmit power, and communication scheduling. The resulting optimization problem is much  more challenging to solve. Nevertheless,  we decompose the  problem into three sub-problems: communication scheduling with fixed transmit power and UAV trajectory sub-problem, UAV trajectory  with fixed transmit power and  communication scheduling  sub-problem, and  transmit power with fixed  UAV trajectory and communication scheduling sub-problem. A three-layer iterative algorithm is then proposed to alternately optimize the communication scheduling, UAV trajectory, and transmit power  based on the SCA method.

\item To demonstrate  our designs more clearly, we consider two simulation scenarios. In the first scenario,  one  UAV-BS collects data from one SN and one  UAV-AP transmits its own data to one AP. In the second  scenario, the UAV-BS and UAV-AP simultaneously serve multiple SNs and APs.  The impact of the weighting factors, UAV trajectory, and transmit power on  the system performance are also studied  to reveal useful insights. Numerical results show that our proposed scheme achieves significantly higher system throughput compared with other benchmarks.
\end{itemize}
The rest of the paper is organized as follows. In Section II, we introduce our system model  and formulate the system throughput maximization  problem. Section III studies  the optimal communication design problem. Section IV investigates the joint 3D UAV trajectory and communication design problem. In Section V,  numerical results are presented to illustrate the  superiority of our scheme. Finally, Section VI  concludes the paper.


\section{System Model}
We consider an integrated network which consists of a UAV-AP and a UAV-BS  based network,  as shown in Fig.~\ref{fig1}. Without loss of generality, we assume that there are $K$ SNs and $L$ APs, which are in   fixed locations.  The SN and AP sets are respectively denoted as $\cal K$ and $\cal L$. The horizontal coordinates of the $k$th SN and $l$th AP are respectively denoted as ${\bf w}_{bk},k \in \cal K $ and ${\bf w}_{ul},l\in \cal L$. We assume that the  UAVs  can adjust their heading as needed. The period $T$ is equally divided  into $N$ time slots indexed by $n=1,...,N$, with duration $\delta$, i.e., $\delta  = \frac{T}{N}$. Note that the duration $\delta$ should be chosen to be sufficiently small so that the UAV's location can be considered unchanged within each time slot even at the maximum flying speed.
As a result, the 3D UAV-AP location at any time slot $n$ is denoted by ${{\bf{w}}_u}\left[ n \right] = \left[ {{{\bf q}_u}\left[ n \right]{\kern 1pt} {\kern 1pt} {\kern 1pt} {\kern 1pt} {H_u}\left[ n \right]} \right]$, where ${\bf q}_u[n]$ and $H_u[n]$ denote the horizontal UAV-AP location and altitude, respectively. Similarly, the 3D UAV-BS location at any time slot $n$ is denoted by ${{\bf{w}}_b}\left[ n \right] = \left[ {{{\bf{q}}_b}\left[ n \right]{\kern 1pt} {\kern 1pt} {\kern 1pt} {\kern 1pt} {H_b}\left[ n \right]} \right]$, where ${\bf q}_b[n]$ and $H_u[n]$ denote the horizontal UAV-BS  location and altitude, respectively.

For the  UAV-to-ground (U2G) channel, ground-to-UAV (G2U) channel, and  UAV-to-UAV (U2U) channel, the  UAV is likely to  establish  LoS links for all U2G, G2U, and U2U channels as reported in \cite{zeng2019accessing}, \cite{khuwaja2018asurvey}. To capture the large-scale fading as well as small-scale fading,   we model  the  U2G, G2U, and  U2U channels as  Rician models \cite{zhan2018energy}, \cite{you20193d},  \cite{zhan2020aerial}\footnote{Although Nakagmi-m also captures the LoS propagation \cite{Azari2019UAV}, it is so sophisticated for analysis in our considered scenarios. To facilitate the system design, we consider a  Rician fading model which is relative simple but still appealing  in practice.}  Thus, the U2G channel coefficient, i.e., UAV-AP to  AP $l$, $\forall l$, at time slot $n$  can be expressed as
\begin{align}
{\bar g_l}\left[ n \right] = \sqrt {{\beta _{a,l}}\left[ n \right]} {\hat g_l}\left[ n \right],
\end{align}
where ${{\beta _{a,l}}\left[ n \right]}$ represents  its  distance-dependent path-loss at time slot $n$, and  ${\hat g_l}\left[ n \right]$ is a complex-valued random variable that denotes the small-scale fading at time slot $n$. Specifically, ${{\beta _{a,l}}\left[ n \right]}$ can be written as
\begin{align}
{\beta _{a,l}}\left[ n \right] = {{{\beta _0}} \over {{{\left( {{{\left\| {{{\bf{q}}_u}\left[ n \right] - {{\bf{w}}_{ul}}} \right\|}^2} + H_u^2\left[ n \right]} \right)}^{{\kappa_a  \mathord{\left/
 {\vphantom {\kappa  2}} \right.
 \kern-\nulldelimiterspace} 2}}}}},
\end{align}
where $\beta_0$ denotes the channel power gain at the reference distance of 1 meter, and $\kappa_a$ denotes the U2G path loss exponent. The small-scale fading, ${{\hat g}_l}\left[ n \right]$, can be modeled as below
\begin{align}
{{\hat g}_l}\left[ n \right] = \sqrt {{{{K_a}} \over {{K_a} + 1}}} {{\tilde g}_l}\left[ n \right] + \sqrt {{1 \over {{K_a} + 1}}} {{\tilde {\tilde g}}_l}\left[ n \right],
\end{align}
where ${{\tilde g}_l}\left[ n \right]$ denotes the deterministic LoS channel component with $\left| {{{\tilde g}_l}\left[ n \right]} \right| = 1$, ${{\tilde {\tilde g}}_l}\left[ n \right] \sim {\cal CN}\left( {0,1} \right)$ denotes the small fading coefficient, and ${{K_a}}$ is the  Rician factor for the  U2G channel.

Similarly, the channel coefficient from    SN $k$ to UAV-BS   at time slot $n$ can be expressed as
\begin{align}
{\bar h_k}\left[ n \right] = \sqrt {{\beta _{s,k}}\left[ n \right]} {\hat h_k}\left[ n \right],
\end{align}
where ${\beta _{s,k}}\left[ n \right] = {{{\beta _0}} \over {{{\left( {{{\left\| {{{\bf{q}}_b}\left[ n \right] - {{\bf{w}}_{bk}}} \right\|}^2} + H_b^2\left[ n \right]} \right)}^{{{{\kappa_s}} \mathord{\left/
 {\vphantom {{{\kappa _s}} 2}} \right.
 \kern-\nulldelimiterspace} 2}}}}}$, ${{\hat h}_k}\left[ n \right] = \sqrt {{{{K_s}} \over {{K_s} + 1}}} {{\tilde h}_k}\left[ n \right] + \sqrt {{1 \over {{K_s} + 1}}} {{\tilde {\tilde h}}_k}\left[ n \right]$,  $\left| {{{\tilde h}_k}\left[ n \right]} \right| = 1$, ${{\tilde {\tilde h}}_k}\left[ n \right] \sim {\cal CN}\left( {0,1} \right)$, and  $\kappa_s$ and $K_s$ denote the G2U path loss exponent and Rician factor, respectively.

 Furthermore, the channel  coefficient for the U2U channel, i.e., UAV-AP  to UAV-BS channel, is given by
\begin{align}
{\bar f}\left[ n \right] = \sqrt {{\beta _{u}}\left[ n \right]} {\hat f}\left[ n \right],
\end{align}
where ${\beta _u}\left[ n \right] = {{{\beta _0}} \over {{{\left( {\left\| {{{\bf{w}}_u}\left[ n \right] - {{\bf{w}}_b}\left[ n \right]} \right\|} \right)}^{{{{\kappa _u}} \mathord{\left/
 {\vphantom {{{\kappa _u}} 2}} \right.
 \kern-\nulldelimiterspace} 2}}}}}$,   ${{\hat f}}\left[ n \right] = \sqrt {{{{K_u}} \over {{K_u} + 1}}} {{\tilde f}}\left[ n \right] + \sqrt {{1 \over {{K_u} + 1}}} {{\tilde {\tilde f}}}\left[ n \right]$, and $\left| {{{\tilde f}}\left[ n \right]} \right| = 1$, ${{\tilde {\tilde f}}}\left[ n \right] \sim {\cal CN}\left( {0,1} \right)$, and  $\kappa_u$ and $K_u$ denote the U2U path loss exponent and Rician factor, respectively.

Note that the path loss exponents for all the channels depend on the environment. For example, it was   shown  in  \cite{ahmed2016on}  that the path loss exponents for U2G/G2U and U2U are $2.32$/$2.51$ and $2.05$, respectively, and the results in \cite{Allred2007SensorFlock} shown that the path loss exponents for U2G and U2U  are $2.13$ and $1.92$, respectively.  Therefore, in the sequel, we set path loss exponents as  ${\kappa _a} = {\kappa _s} = {\kappa _u} = 2$  that are  consistent with the most existing works \cite{xu2018uav,lyu2018uav,zhou2018computation,power2018meng,wu2018capacity}. Although we assume that the path loss exponent is $2$, it can be easily extended to the other cases.
In addition, we assume that   the Rician factors ${{K_a}}$, ${{K_s}}$, and ${{K_u}}$ are  all   invariant over time slot $n$ by considering the  following reasons. First, since our  scenario is considered in the rural and/or suburban district, i.e., clear airspace, the Rician factor thus can be approximately treated to be independent of the varying UAV locations. Second, especially for the long period flying time $T$, the most time for UAV is to stay stationary above the ground nodes, and thus can be considered as invariant at most of the time.

In addition, for the ground-to-ground (G2G) channel, we assume that the G2G channel follows  Rayleigh fading due to  the rich scattering in the environment. Therefore,   the channel coefficient  from the $k$th SN to the $l$th AP  can be expressed as
\begin{align}
{{\bar h}_{k,l}} = \sqrt {{{\tilde h}_{k,l}}} {{\hat h}_{k,l}},
\end{align}
where ${{\tilde h}_{k,l}} = {{{\beta _0}} \over {{{\left\| {{{\bf{w}}_{bk}} - {{\bf{w}}_{ul}}} \right\|}^\alpha }}}$ stands for the large-scale path loss, $\alpha$ represents the G2G path loss exponent, and ${{\hat h}_{k,l}} \sim {\cal CN}\left( {0,1} \right)$ denotes the small-scale fading.

To facilitate the system design, we assume the  widely used wake-up communication scheduling approach \cite{wu2018Joint},\cite{hua2019energy}, and \cite{power2018meng}, where the UAV-BS (UAV-AP) can only communicate with at most one SN (AP) in  any time slot $n$. Define the indicator variable $y_k[n],\forall k,n$  and $x_l[n],\forall l,n$ for the UAV-BS and UAV-AP based network, respectively. The UAV-BS serves the $k$th SN if $y_k[n]=1$, otherwise, $y_k[n]=0$. Similarly, if $x_l[n]=1$, the UAV-AP migrates   the  data to the $l$th AP, and  no data is transmitted if $x_l[n]=0$. Thus, we have the  following scheduling constraints
\begin{align}
&\sum\nolimits_{l = 1}^L {{x_l}\left[ n \right]}  \le 1, ~{x_l}\left[ n \right] \in \left\{ {0,1} \right\},\forall l,n,\label{systemmodel4} \\
&\sum\nolimits_{k = 1}^K {{y_k}\left[ n \right]}  \le 1,~{y_k}\left[ n \right] \in \left\{ {0,1} \right\}, \forall k,n.
\end{align}
If the $l$th AP is awakened to communicate with the UAV-AP  at time slot $n$, the achievable downlink rate  of   the  $l$th AP is given by
\begin{align}
{\bar R}_l^u\left[ n \right] = {\log _2}\left( {1 + {{{{\left| {{{\bar g}_l}[n]} \right|}^2}{p^u}\left[ n \right]} \over {\sum\nolimits_{k = 1}^K {{{\left| {{{\bar h}_{k,l}}} \right|}^2}{y_k}\left[ n \right]p_k^s\left[ n \right] + {\sigma ^2}} }}} \right), \label{systemmodel5}
\end{align}
where $p^u[n]$ and $p_k^s[n]$ respectively denote the UAV-AP and $k$th SN transmit power  at time slot $n$, and $\sigma^2$ represents the noise power at the receiver.

When $y_k[n]=1$, the uplink transmission rate of SN $k$ is given by
\begin{align}
{\bar R}_k^s\left[ n \right] = {\log _2}\left( {1 + {{{{\left| {{{\bar h}_k}\left[ n \right]} \right|}^2}p_k^s\left[ n \right]} \over {\sum\nolimits_{l = 1}^L {{{\left| {{\bar f}\left[ n \right]} \right|}^2}{x_l}\left[ n \right]{p^u}\left[ n \right]}  + {\sigma ^2}}}} \right). \label{systemmodel6}
\end{align}
Obviously, \eqref{systemmodel6} can be simplified as
\begin{align}
{\bar R}_k^s\left[ n \right] = {\log _2}\left( {1 + {{{{\left| {{{\bar h}_k}\left[ n \right]} \right|}^2}p_k^s\left[ n \right]} \over {{{\left| {{\bar f}\left[ n \right]} \right|}^2}{p^u}\left[ n \right] + {\sigma ^2}}}} \right).
\end{align}
This is because with \eqref{systemmodel4},  if   AP $l$ is communicated with UAV-AP at time  slot $n$, namely ${x_l}\left[ n \right] = 1$,  we have $\sum\nolimits_{l = 1}^L {{{\left| {{\bar f}\left[ n \right]} \right|}^2}{x_l}\left[ n \right]{p^u}\left[ n \right]}  = {\left| {{\bar f}\left[ n \right]} \right|^2}{p^u}\left[ n \right]$; if no AP is activated,  the  transmission power of UAV-AP $p_u[n]$ must be zero.

Note that since the channels $\left\{ {{{\bar h}_{k,l}},{{\bar h}_k}\left[ n \right],{{\bar g}_l}\left[ n \right],\bar f\left[ n \right]} \right\}$ are the random variables, ${\bar R}_k^s\left[ n \right]$ and ${\bar R}_l^u\left[ n \right]$ are also the random variables.  Additionally, since the probability distribution of ${\bar R}_k^s\left[ n \right]$ and ${\bar R}_l^u\left[ n \right]$ are challenging to obtain, we are interested in the expected/average achievable rate, defined as $R_k^s\left[ n \right] = {\mathbb E}\left\{ {\bar R_k^s\left[ n \right]} \right\}$ and $R_l^u\left[ n \right] = {\mathbb E}\left\{ {\bar R_l^u\left[ n \right]} \right\}$. However, the closed-form expressions of $R_k^s\left[ n \right]$ and $R_l^u\left[ n \right]$  are unsolvable due to the difficulty
of deriving its probability distribution. To address this issue, we obtain their  approximation results based on the following theorem.
\begin{theorem} \label{theorem1}
If $X$ is a non-negative positive random variable and $Y$ is a positive random variable, and $X$ and $Y$ are independent, the following approximation result holds
\begin{align}
{\mathbb E}\left\{ {{{\log }_2}\left( {1 + {X \over Y}} \right)} \right\} \approx {\log _2}\left( {1 + {{{\mathbb E}\left\{ X \right\}} \over {{\mathbb E}\left\{ Y \right\}}}} \right). \label{theorem1result1}
\end{align}
\end{theorem}
\begin{IEEEproof}
Please refer to Appendix~\ref{appendix1}.
\end{IEEEproof}
Based on Theorem~\ref{theorem1}, we can, respectively, recast $R_k^s\left[ n \right]$ and $R_l^u\left[ n \right]$ as
\begin{align}
R_k^s\left[ n \right] &\approx {\log _2}\left( {1 + {{\mathbb E\left\{ {{{\left| {{{\bar h}_k}\left[ n \right]} \right|}^2}} \right\}p_k^s\left[ n \right]} \over {\mathbb E\left\{ {{{\left| {\bar f\left[ n \right]} \right|}^2}} \right\}{p^u}\left[ n \right] + {\sigma ^2}}}} \right)\notag\\
&\overset{\triangle}= \log_2\left( {1 + {{{h_k}\left[ n \right]p_k^s\left[ n \right]} \over {f\left[ n \right]{p^u}\left[ n \right] + {\sigma ^2}}}} \right) \label{ergodicR_ks}
\end{align}
and
\begin{align}
R_l^u\left[ n \right] &\approx {\log _2}\left( {1 + {{\mathbb E\left\{ {{{\left| {{{\bar g}_l}[n]} \right|}^2}} \right\}{p^u}\left[ n \right]} \over {\sum\nolimits_{k = 1}^K {\mathbb E\left\{ {{{\left| {{{\bar h}_{k,l}}} \right|}^2}} \right\}{y_k}\left[ n \right]p_k^s\left[ n \right] + {\sigma ^2}} }}} \right)\notag\\
&\overset{\triangle}= {\log _2}\left( {1 + {{{g_l}[n]{p^u}\left[ n \right]} \over {\sum\nolimits_{k = 1}^K {{{\tilde h}_{k,l}}{y_k}\left[ n \right]p_k^s\left[ n \right] + {\sigma ^2}} }}} \right), \label{ergodicR_ul}
\end{align}
where ${h_k}\left[ n \right] = \beta _{s,k}\left[ n \right],{g_l}\left[ n \right] = \beta _{a,l}\left[ n \right]$, and $f\left[ n \right] = \beta _u\left[ n \right]$. The accuracy for the approximation results of $R_k^s\left[ n \right]$ and $R_l^u\left[ n \right]$ will be   evaluated later in Section V.

In this paper, we focus on the joint  design of the UAV trajectory, communication scheduling, and transmit power  to maximize the integrated network throughput, i.e., the sum throughput of the  UAV-BS and UAV-AP based networks. Define sets $A = \left\{ {{x_l}\left[ n \right],{y_k}\left[ n \right],\forall l,k,n} \right\}$,  $P = \left\{ {{p^u}\left[ n \right],p_k^s\left[ n \right],\forall k,n} \right\}$, and $Q = \left\{ {{{\bf{w}}_u}\left[ n \right],{{\bf{w}}_b}\left[ n \right],\forall n} \right\}$. Then, the problem can be formulated as\footnote{The formulated problem can be easily extended to the  case where the different links have different priorities by setting different weighting factors on the different links in the objective function.}
\begin{subequations} \label{P1}
\begin{align}
&\mathop {\max }\limits_{A,P,Q} {\beta _1}\sum\limits_{n = 1}^N {\sum\limits_{k = 1}^K {{y_k}\left[ n \right]} } {\log _2}\left( {1 + \frac{{{h_k}\left[ n \right]p_k^s\left[ n \right]}}{{f\left[ n \right]{p^u}\left[ n \right] + {\sigma ^2}}}} \right) + \notag\\
&{\beta _2}\sum\limits_{n = 1}^N {\sum\limits_{l = 1}^L x_l[n]{{{\log }_2}\left( {1 + \frac{{{g_l}[n]{p^u}\left[ n \right]}}{{\sum\nolimits_{k = 1}^K {{{\tilde h}_{k,l}}{y_k}\left[ n \right]p_k^s\left[ n \right] + {\sigma ^2}} }}} \right)} } \label{P1_const0}\\
{\rm s.t.}~&\sum\nolimits_{l = 1}^L {{x_l}\left[ n \right]}  \le 1, ~{x_l}\left[ n \right] \in \left\{ {0,1} \right\},\forall l,n,\label{P1_const1}\\
&\sum\nolimits_{k = 1}^K {{y_k}\left[ n \right]}  \le 1,~{y_k}\left[ n \right] \in \left\{ {0,1} \right\}, \forall k,n,\label{P1_const2}\\
&0 \le {p^u}\left[ n \right] \le p_{\max }^u,\forall n,\label{P1_const3}\\
&0 \le p_k^s\left[ n \right] \le p_{\max }^s,\forall k,n,\label{P1_const4}\\
&\left\| {{H_i}\left[ n \right] - {H_i}\left[ {n - 1} \right]} \right\| \le {V_z}\delta ,\forall n,i \in \left\{ {b,u} \right\},\label{P1_const5}\\
&{H_{\min }} \le {H_i}\left[ n \right] \le {H_{\max }},\forall n,i \in \left\{ {b,u} \right\},\label{P1_const6}\\
&{H_i}\left[ 0 \right] = {H_{{I_i}}},{H_i}\left[ N \right] = {H_{{F_i}}},i \in \left\{ {b,u} \right\},\label{P1_const7}\\
&\left\| {{{\bf{q}}_i}\left[ n \right] - {{\bf{q}}_i}\left[ {n - 1} \right]} \right\| \le {V_{xy}}\delta ,\forall n,i \in \left\{ {b,u} \right\},\label{P1_const8}\\
&{{\bf{q}}_i}\left[ 0 \right] = {{\bf{q}}_{{I_i}}},{{\bf{q}}_i}\left[ N \right] = {{\bf{q}}_{{F_i}}},i \in \left\{ {b,u} \right\},\label{P1_const9}\\
&{\left\| {{{\bf{q}}_b}\left[ n \right] - {{\bf{q}}_u}\left[ {n - 1} \right]} \right\|^2} + {\left\| {{H_b}\left[ n \right] - {H_u}\left[ n \right]} \right\|^2} \ge {d_{\min }^2},\forall n,\label{P1_const10}
\end{align}
\end{subequations}
where $\beta_1$ and $\beta_2$ are the  weighting factors.  Equations \eqref{P1_const3} and \eqref{P1_const4} represent  the transmit power constraints, with $p_{\max }^u$ and $p_{\max }^s$  denoting the maximum power limits at the UAV-AP and SNs, respectively. Equations \eqref{P1_const5}\text{-}\eqref{P1_const9} denotes the UAV trajectory constraints, where $V_z$ and $V_{xy}$ respectively  denote the maximum UAV vertical  and horizontal speed, $H_{I_i}$ and ${\bf q}_{I_i}$ represent the initial location for UAV $i$, $H_{F_i}$ and ${\bf q}_{F_i}$  represents the final location for UAV $i$. Finally, \eqref{P1_const10} denotes the  collision avoidance constraint between the two UAVs with a minimum safety distance $d_{\rm min}$.


\section{Globally optimal communication design }
In this section, we  obtain the globally optimal solution to \eqref{P1} for the particular case when the two UAV trajectories  are pre-determined. In practice, for a large number of UAV applications, the flight paths are fixed, e.g., the UAV flies in a circular path along the cell edge to serve the cell-edge users, or the UAV flies in a straight line to communicate with the ground users  \cite{lyu2018uav},\cite{lyu2016cyclical}. As  a result, \eqref{P1} is simplified as
\begin{subequations} \label{P2}
\begin{align}
&\mathop {\max }\limits_{A,P} {\beta _1}\sum\limits_{n = 1}^N {\sum\limits_{k = 1}^K {{y_k}\left[ n \right]} } {\log _2}\left( {1 + \frac{{{h_k}\left[ n \right]p_k^s\left[ n \right]}}{{f\left[ n \right]{p^u}\left[ n \right] + {\sigma ^2}}}} \right)\notag\\
& + {\beta _2}\sum\limits_{n = 1}^N {\sum\limits_{l = 1}^L x_l[n]{{{\log }_2}\left( {1 + \frac{{{g_l}[n]{p^u}\left[ n \right]}}{{\sum\nolimits_{k = 1}^K {{{\tilde h}_{k,l}}{y_k}\left[ n \right]p_k^s\left[ n \right] + {\sigma ^2}} }}} \right)} } \label{P2_const0}\\
&\quad{\rm s.t.}~\eqref{P1_const1}\text{-}\eqref{P1_const4}.
\end{align}
\end{subequations}
Problem \eqref{P2} is difficult to solve due to the coupled power and communication scheduling in \eqref{P2_const0} and the binary variables in \eqref{P1_const1} and \eqref{P1_const2}.  However, we show how to optimally solve \eqref{P2}  by using monotonic optimization theory \cite{tuy2000monotonic},\cite{zhang2013monotonic}. First, it is observed that $y_k[n]$ and $x_l[n]$ in \eqref{P2_const0} can be moved into the numerator of the logarithm terms since  $y_k[n]=1$ for at most  one $k$ ($x_l[n]=1$ for at most  one $l$). Either way, the terms where $y_k[n]=0$  and  $x_l[n]=0$ do not contribute to the objective valuable. Defining $\tilde p_l^u\left[ n \right] = {p^u}\left[ n \right]{x_l}\left[ n \right]$ for all $l$, $\tilde p_k^s\left[ n \right] = p_k^s\left[ n \right]{y_k}\left[ n \right]$ for all $k$, and   $\tilde P = \left\{ {{{\tilde p}^u}\left[ n \right],\tilde p_k^s\left[ n \right],\forall k,n} \right\}$, we formulate the following problem:
\begin{small}
\begin{subequations} \label{P3}
\begin{align}
&\mathop {\max }\limits_{\tilde P} {\beta _1}\sum\limits_{n = 1}^N {\sum\limits_{k = 1}^K {{{\log }_2}\left( {1 + \frac{{{h_k}\left[ n \right]\tilde p_k^s\left[ n \right]}}{{M\sum\nolimits_{i \ne k}^K {\tilde p_i^s\left[ n \right]}  + \sum\nolimits_{l = 1}^L {f\left[ n \right]\tilde p_l^u\left[ n \right]}  + {\sigma ^2}}}} \right)} }    \notag\\
&+{\mkern 1mu} {\beta _2}\sum\limits_{n = 1}^N {\sum\limits_{l = 1}^L {{\log }_2}\left( {1 + \frac{{{g_l}[n]\tilde p_l^u\left[ n \right]}}{{M\sum\nolimits_{i \ne l}^L {\tilde p_i^u\left[ n \right] + } \sum\nolimits_{k = 1}^K {{{\tilde h}_{k,l}}\tilde p_k^s\left[ n \right] + {\sigma ^2}} }}} \right)}\label{P3_const0}\\
&\quad {\rm s.t.}~{\tilde P}\in{\cal { P}},
\end{align}
\end{subequations}
\end{small}
where ${\cal {P}}=\left\{ {\tilde P|0 \le \tilde p_l^u\left[ n \right] \le p_{\max }^u,0 \le \tilde p_k^s\left[ n \right] \le p_{\max }^s,\forall k,l,n} \right\}$, and $M$ is a sufficiently large  penalty factor.

\begin{theorem} \label{theorem2}
Problem \eqref{P3} is equivalent to \eqref{P2}.
\end{theorem}
\begin{IEEEproof}
Please refer to Appendix~\ref{appendix2}.
\end{IEEEproof}
There is no standard method  to  obtain the  optimal solution to \eqref{P3} due to the coupled transmit power  in the objective function. However, by exploiting the hidden monotonicity in the problem,  we obtain the optimal solution to problem  \eqref{P3} by following two steps. We first  transform problem \eqref{P3} into an equivalent canonical monotonic optimization formulation. Then, we apply a sequence of ployblocks to approach the optimal vertex using POA method. Specifically, by introducing the auxiliary variables $\chi _k[n]$ and $\bar \chi _l[n]$, problem \eqref{P3} can be equivalently written as
\begin{subequations} \label{P4}
\begin{align}
&\mathop {\max }\limits_{{\chi _k}\left[ n \right],{{\bar \chi }_l}\left[ n \right]} {\beta _1}\sum\limits_{n = 1}^N {\sum\limits_{k = 1}^K {{{\log }_2}\left( {1 + {\chi _k}\left[ n \right]} \right)} } +  {\beta _2}\sum\limits_{n = 1}^N {\sum\limits_{l = 1}^L {{{\log }_2}\left( {1 + {{\bar \chi }_l}\left[ n \right]} \right)} }\\
&\quad{\rm s.t.}~\left( {{{\bm \chi} _k}\left[ n \right],{{\bm  {\bar  \chi }}_l}\left[ n \right]} \right) \in {\cal G},
\end{align}
\end{subequations}
where ${{\bm{\chi }}_k}\left[ n \right]$ and ${{{{\bm{\bar \chi }}}_l}\left[ n \right]}$ are the collections of  $\chi _k[n]$ and $\bar \chi _l[n]$, respectively, and  the normal set ${\cal G}$ is defined in \eqref{P4_const1}. Note that the signal-to-interference-plus-noise-ratio (SINR) for the UAV-BS and UAV-AP based networks  must be non-negative. Therefore, both $\chi _k[n]$ and $\bar \chi _l[n]$ must be  no smaller than  than zero, i.e., ${\cal H}{\rm{ = }}\left\{ {\left( {{\chi _k}[n],{{\bar \chi }_l}[n]} \right)|{\chi _k}[n] \ge 0,{{\bar \chi }_l}[n] \ge 0,\forall k,l,n} \right\}$. It can be seen that
the objective function in \eqref{P4} is an   increasing function with  $\chi _k[n]$ and $\bar \chi _l[n]$. In addition,  the power allocation  $\tilde p_l^u[n]$ and $\tilde p_k^s[n]$ in   the normal $\cal G$  in \eqref{P4_const1} can be optimally obtained  when  $\chi _k[n]$ and $\bar \chi _l[n]$ are fixed \big(see \eqref{P5} for more  details\big). Therefore, \eqref{P4} is in the canonical form of a monotonic optimization problem, and the optimal solution can be   obtained  by searching  the upper boundary of the feasible set  using the POA method, which is summarized in Algorithm~\ref{alg1} \cite{tuy2000monotonic},\cite{zhang2013monotonic}.
\newcounter{mytempeqncnt0}
\begin{figure*}
\normalsize
\setcounter{mytempeqncnt0}{\value{equation}}
\begin{align}
&{\cal G}= {\rm{ }}\left\{ {\left( {{\chi _k}\left[ n \right],{{\bar \chi }_l}\left[ n \right]} \right){\rm{|}}{\chi _k}\left[ n \right] \le {{{h_k}\left[ n \right]\tilde p_k^s\left[ n \right]} \over {M\sum\nolimits_{i \ne k}^K {\tilde p_i^s\left[ n \right]}  + \sum\nolimits_{l = 1}^L f \left[ n \right]\tilde p_l^u\left[ n \right] + {\sigma ^2}}}} \right.,\notag\\
&\qquad\qquad\qquad\qquad\qquad\qquad\left. {{{\bar \chi }_l}\left[ n \right] \le {{{g_l}[n]\tilde p_l^u\left[ n \right]} \over {M\sum\nolimits_{i \ne l}^L {\tilde p_i^u\left[ n \right] + } \sum\nolimits_{k = 1}^K {{{\tilde h}_{k,l}}\tilde p_k^s\left[ n \right] + {\sigma ^2}} }},\forall l,k,n,\tilde P \in P} \right\}.\label{P4_const1}
%
\end{align}
\hrulefill 
\vspace*{4pt} 
\end{figure*}
\begin{algorithm}[htbp]
\caption{ Polyblock Outer Approximation  (POA) based method} \label{alg1}
\begin{algorithmic}[1]
\STATE  \textbf{Initialize} polyblock ${\cal S}^1$ with vertex ${\bm v}^1=({{\bm{\chi }}_k^1}\left[ n \right],{{{{\bm{\bar \chi }}}_l^1}\left[ n \right]})$, where $\chi _k^1[n] = \frac{{{h_k}\left[ n \right]p_{\max }^s}}{{{\sigma ^2}}}$ and $\bar \chi _l^1[n]=\frac{{{g_l}\left[ n \right]p_{\max }^u}}{{{\sigma ^2}}}$ for $\forall k,l,n$;  ${\cal T}^1=\{{\bm v}^1\}$,  maximum tolerance $\epsilon=10^{-2}$, and  iterative index $t=1$. \\
\STATE  \textbf{Repeat}
\STATE \quad Compute  the projection of ${{{\bf{v}}^t}}$ on the upper  boundary of \\
\quad ${\cal G}$, denoted as ${{\bm \pi} ^{\cal G}}\left( {{{\bf{v}}^t}} \right)$, via   Algorithm~\ref{alg2}.
\STATE \quad With ${{\bm \pi} ^{\cal G}}\left( {{{\bf{v}}^t}} \right)$,  generate $M$ new vertices $\left\{ {{\bf{\tilde v}}_1^t,...,{\bf{\tilde v}}_M^t} \right\}$,\\
\quad  where ${\bf{\tilde v}}_i^t = {{\bf{v}}^t} - \left( {v_i^t - {\bm \pi} _i^{\cal G}\left( {{{\bf{v}}^t}} \right)} \right){{\bf{e}}_i}$  for $i=1,...,M$.
\STATE \quad Construct a smaller polyblock ${\cal S}^{t+1}$ with  vertex set  \\
\quad ${\cal T}^{t+1}$ by replacing  ${{{\bf{v}}^t}}$ in ${\cal T}^{t}$  with $M$  new vertices \\
\quad $\left\{ {{\bf{\tilde v}}_1^t,...,{\bf{\tilde v}}_M^t} \right\}$.\\
\STATE \quad Find ${{{\bf{v}}^{t+1}}}$ as the candidate vertex that  maximizes the \\
\quad objective function of problem \eqref{P4}  over set ${\cal T}^{t+1} \cap {\cal H} $.
\STATE \quad $t=t+1$.
\STATE   \textbf{Until} $\mathop {\max }\limits_i \left\{ {\frac{{\left\| {{\bf{v}}_i^t - {\bm \pi} _i^{\cal G}\left( {{{\bf{v}}^t}} \right)} \right\|}}{{\left\| {{\bf{v}}_i^t} \right\|}}} \right\} \le \epsilon$.\\
\STATE Output  optimal transmit power  $\{{\tilde p_k^{s,*}\left[ n \right]}\}$ and $\{{\tilde p_l^{u,*}\left[ n \right]}\}$  by  computing
${{\bm \pi} ^{\cal G}}\left( {{{\bf{v}}^t}} \right)$ in Algorithm~\ref{alg2}.
\end{algorithmic}
\end{algorithm}
\begin{figure}[htbp]
\centerline{\includegraphics[width=3in]{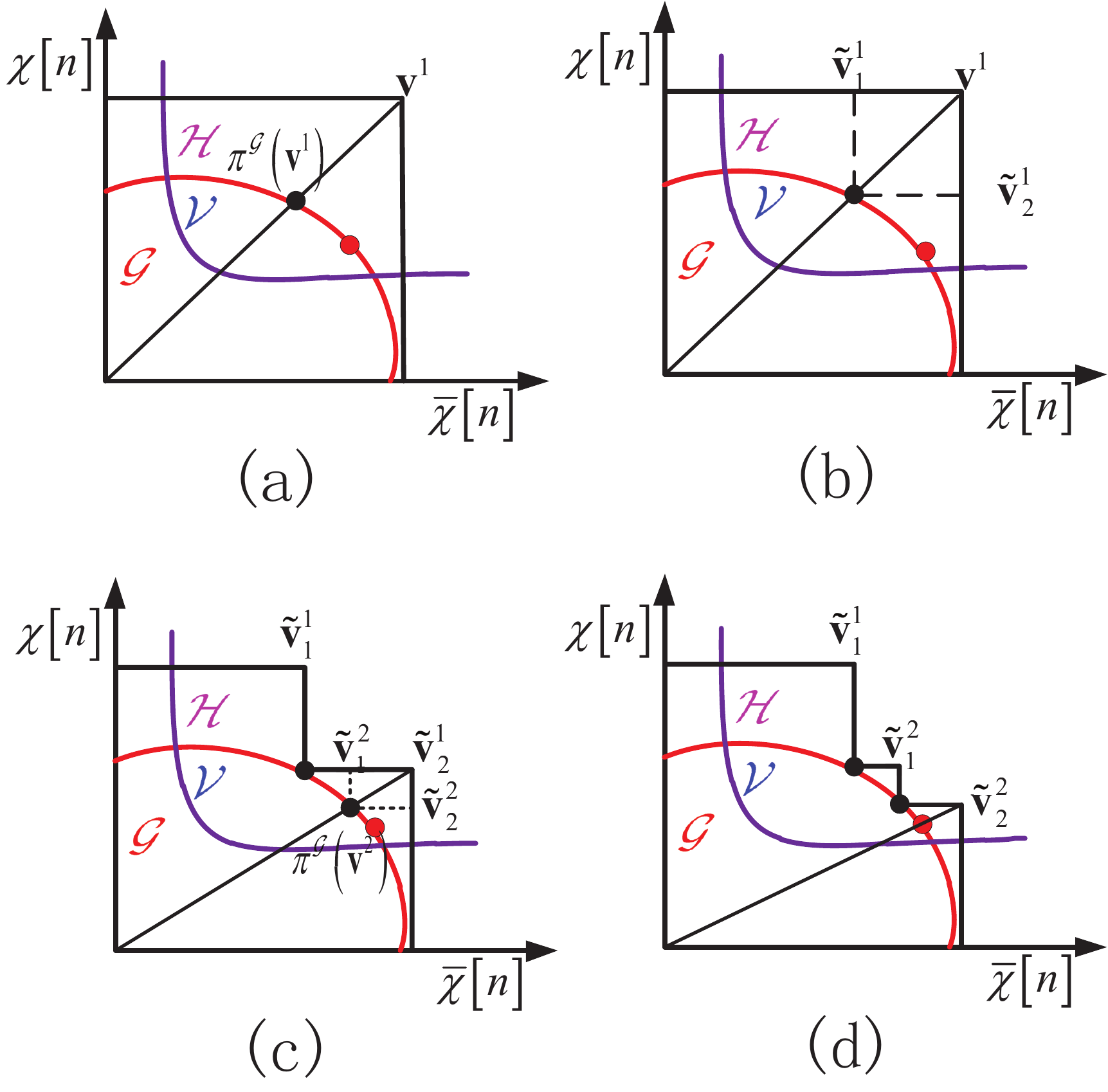}}
\caption{Illustration of  Algorithm~\ref{alg1}, where the red circle denotes the optimal point on the upper boundary of the feasible set  $\cal V{\rm{ = }}\cal G \cap \cal H$.} \label{alg1fig}
\end{figure}

To explain  Algorithm~\ref{alg1} more clearly, we provide a simple case that includes  two optimization variables $\chi[n]$ and $\bar \chi [n]$ as shown in  Fig.~\ref{alg1fig}. In the initial stage of Algorithm~\ref{alg1}, we set $\chi _k^1[n] = \frac{{{h_k}\left[ n \right]p_{\max }^s}}{{{\sigma ^2}}}$ and $\bar \chi _l^1[n]=\frac{{{g_l}\left[ n \right]p_{\max }^u}}{{{\sigma ^2}}}$, $\forall k,l,n$, and define vertex ${\bf v}^1$ as ${\bf v}^1=({{\bm{\chi }}_k^1}\left[ n \right],{{{{\bm{\bar \chi }}}_l^1}\left[ n \right]})$.  It is clear that polyblock ${\cal S}^1$ is a box $\left[ {{\bf{0}}~{{\bf{v}}^1}} \right]$ comprising the feasible set $\cal V{\rm{ = }}\cal G \cap \cal H$. In step 3, we calculate the projection of vertex ${\bf v}^1$ onto set $\cal G$, i.e., ${{\bm \pi} ^{\cal G}}\left( {{{\bf{v}}^1}} \right)$ \big(see Fig.~\ref{alg1fig} (a)\big). In step 4, based on vertex ${\bf v}^1$ and ${{\bm \pi} ^{\cal G}}\left( {{{\bf{v}}^1}} \right)$, we generate $M=(K+L)N$ new vertices, denoted as $\left\{ {{\bf{\tilde v}}_1^1,...,{\bf{\tilde v}}_M^1} \right\}$, where ${\bf{\tilde v}}_i^1 = {{\bf{v}}^1} - \left( {v_i^1 - {\bm \pi} _i^{\cal G}\left( {{{\bf{v}}^1}} \right)} \right){{\bf{e}}_i}$ \big(see Fig.~\ref{alg1fig} (b)\big). Here, $v_i^1$ is the $i$th element of ${\bf v}^1$, ${{\bm \pi}_i ^{\cal G}}\left( {{{\bf{v}}^1}} \right)$ is the $i$th element of ${{\bm \pi} ^{\cal G}}\left( {{{\bf{v}}^1}} \right)$, and ${\bm e}_i$ denotes  the $i$th column of the identity matrix. In step 5, we shrink the polyblock ${\cal S}^1$, denoted as polyblock ${\cal S}^2$,  by replacing ${\bm v}^1$ with the new vertices set  ${\cal T}^{2}$, i.e., ${{\cal T}^2} = \left( {{{\cal T}^1} - {{\bf{\tilde v}}^1}} \right) \cup \left\{ {{\bf{\tilde v}}_1^1,...,{\bf{\tilde v}}_M^1} \right\}$ \big(see Fig.~\ref{alg1fig} (c)\big). It can be observed that  polyblock ${\cal S}^2$ still contains the feasible set  $\cal V$ but is smaller than polyblock ${\cal S}^1$. Then, we choose the vertex from ${\cal T}^2 \cap H$ that maximizes the  objective function of problem \eqref{P4} \big(see ${\bf{\tilde v}}_2^1$ in  Fig.~\ref{alg1fig} (c)\big). Similarly, we repeat the above procedures to find  a smaller and tighter polyblock that satisfies ${\cal S}^1\supset {\cal S}^2\supset  \cdots  \supset {\cal V}$ \big(see   Fig.~\ref{alg1fig} (d)\big). Therefore, Algorithm~\ref{alg1} will finally approach the optimal solution when $\mathop {\max }\limits_i \left\{ {\frac{{\left\| {{\bf{v}}_i^t - {\bm \pi} _i^{\cal G}\left( {{{\bf{v}}^t}} \right)} \right\|}}{{\left\| {{\bf{v}}_i^t} \right\|}}} \right\} \le \epsilon$.
\begin{algorithm}[htbp]
\caption{Bisection Search to Compute ${{\bm \pi} ^{\cal G}}\left( {{{\bf{v}}^t}} \right)$ } \label{alg2}
\begin{algorithmic}[1]
\STATE  \textbf{Initialize:} $\lambda_{\rm min}=0$, $\lambda_{\rm max}=1$, $\epsilon=10^{-2}$. \\
\STATE  \textbf{Repeat}
\STATE \quad Compute $\lambda  = \frac{{{\lambda _{\min }} + {\lambda _{\max }}}}{2}$\\
\STATE \quad Check the feasibility  of problem \eqref{P5}, i.e., $\lambda {{\bf{v}}^t} \in { \cal G}$. If \\ \quad yes, let ${\lambda _{\min }} = \lambda $, otherwise,  let ${\lambda _{\max }} = \lambda $.
\STATE   \textbf{Until} ${\lambda _{\max }} - {\lambda _{\min }} \le \epsilon $\\
\STATE Output  $\lambda  = {\lambda _{\min }}$ and ${\pi ^G}\left( {{{\bf{v}}^t}} \right) = \lambda {{\bf{v}}^t}$. The optimal power allocation $\{{\tilde p_k^{s,*}\left[ n \right]}\}$ and $\{{\tilde p_l^{u,*}\left[ n \right]}\}$ are  obtained by solving problem \eqref{P5} for $\lambda  = {\lambda _{\min }}$.
\end{algorithmic}
\end{algorithm}
\begin{subequations} \label{P5}
\begin{align}
&{\rm{Find}}{\kern 1pt} {\kern 1pt} {\kern 1pt} {\kern 1pt} {\rm{solutions:}}{\kern 1pt} {\kern 1pt} \{ {\tilde p_k^s\left[ n \right],\tilde p_l^u\left[ n \right]} \}\notag \\
&{\rm s.t.}~\lambda {\chi _k}\left[ n \right] \le \frac{{{h_k}\left[ n \right]\tilde p_k^s\left[ n \right]}}{{M\sum\nolimits_{i \ne k}^K {\tilde p_i^s\left[ n \right]}  + \sum\nolimits_{l=1}^Lf\left[ n \right]{{\tilde p}_l^u}\left[ n \right] + {\sigma ^2}}},\label{P5_const1}\\
&~~~~\lambda {{\bar \chi }_l}\left[ n \right] \le \frac{{{g_l}[n]\tilde p_l^u\left[ n \right]}}{{M\sum\nolimits_{i \ne l}^L {\tilde p_i^u\left[ n \right] + } \sum\nolimits_{k = 1}^K {{{\tilde h}_{k,l}}\tilde p_k^s\left[ n \right] + {\sigma ^2}} }}.\label{P5_const2}\\
&~~~~0 \le {\tilde p_l^u}\left[ n \right] \le p_{\max }^u,\forall n,\label{P5_const3}\\
&~~~~0 \le \tilde p_k^s\left[ n \right] \le p_{\max }^s,\forall k,n.\label{P5_const4}
\end{align}
\end{subequations}
Following   \cite[Proposition 6]{tuy2000monotonic}, the value ${{\bm \pi} ^{\cal G}}\left( {{{\bf{v}}^t}} \right)$ in step 3 of  Algorithm~\ref{alg1} can be calculated as follows: ${\pi ^{\cal G}}\left( {{{\bf{v}}^t}} \right) = \lambda {{\bf{v}}^t}$, where $\lambda  = \max \left\{ {a|a{{\bf{v}}^t} \in {\cal G}} \right\}$, and the details are  summarized in Algorithm~\ref{alg2}. Note that \eqref{P5} in Algorithm~\ref{alg2} can be recast as  a  linear optimization problem by transforming the fractional constraints \eqref{P5_const1} and \eqref{P5_const2} into linear forms, and thus can be optimally solved. Then, we can recover the optimal transmit power for \eqref{P2} using the following  steps: if $\tilde p_k^s\left[ n \right] {\rm{ > }} 0$, $y_k[n]=1$ and $p_k^s\left[ n \right]{\rm{ = }}\tilde p_k^s\left[ n \right]$; and if $\tilde p_k^s\left[ n \right]{\rm{ = }}0$, $y_k[n]=0$ and $p_k^s\left[ n \right]{\rm{ = }}0$. Similar to $ p^u\left[ n \right]$, if $\tilde p_l^u\left[ n \right] {\rm{ > }} 0$, $x_l[n]=1$ and $p^u\left[ n \right]{\rm{ = }}\tilde p_l^u\left[ n \right]$.
\subsection{Optimality and Complexity  Analysis}
 The optimality analysis of  Algorithm~\ref{alg1} is given as follows:
In every iteration in step 4 of Algorithm~\ref{alg1}, we can generate a subsequence thought as the ``off-springs'' through a series of projections. Therefore,  an infinite length of sequences can be obtained as the number of iteration increases. Clearly, we always have  the sequence: ${{\bf{v}}^1} \succeq  {{\bf{v}}^2} \succeq   \cdots  \succeq  {{\bf{v}}^t} \succeq  {{\bf{v}}^{t + 1}} \cdots  \succeq  {\bf{0}}$. Hence, $\mathop {\lim }\limits_{t \to \infty } \left\| {{{\bf{v}}^t} - {{\bf{v}}^{t + 1}}} \right\| \to 0$. We assume that ${{\bf{v}}^{t + 1}}$ is a vector picking from the ``off-springs'' subsequence, i.e., subsequence $\left\{ {{\bf{\tilde v}}_1^t,...,{\bf{\tilde v}}_M^t} \right\}$, obtained from ${{\bf{v}}^t}$. Without loss of generality, we assume that ${{\bf{v}}^{t + 1}} = {\bf{\tilde v}}_m^t$.
As $t \to \infty $, we have $\mathop {\lim }\limits_{t \to \infty } \left\| {{{\bf{v}}^t} - {{\bf{v}}^{t + 1}}} \right\| = \mathop {\lim }\limits_{t \to \infty } \left\| {{{\bf{v}}^t} - {\bf{\tilde v}}_m^t} \right\|\overset{(a)} {=} v_m^t - {\bm \pi} _m^G\left( {{{\bf{v}}^t}} \right) \to 0$, where $(a)$ holds since ${\bf{\tilde v}}_m^t = {{\bf{v}}^t} - \left( {v_m^t - {\bm \pi} _m^{\cal G}\left( {{{\bf{v}}^t}} \right)} \right){{\bf{e}}_m}$.
In addition, since ${\bm \pi} _m^G\left( {{{\bf{v}}^t}} \right) = \lambda v_m^t$, we  have $\left| {v_m^t - \lambda v_m^t} \right| \to 0$, which implies $\lambda\to 1$.  We thus have $\mathop {\lim }\limits_{t \to \infty } \left\| {{{\bf{v}}^t} - {\bm \pi} ^G\left( {{{\bf{v}}^t}} \right)} \right\|=\mathop {\lim }\limits_{t \to \infty } \left\| {{{\bf{v}}^t} - \lambda {{\bf{v}}^t}} \right\| \to 0$. Recall that the optimal solution lies on the upper boundary of $\cal G$, and ${{{\bf{v}}^t}}$ is a maximizer over the  above sequence, the globally optimal solution is thus obtained. The reader can also refer to  \big[Theorem 1, \cite{tuy2000monotonic}\big] for  more details.

The computational complexity   of  Algorithm~\ref{alg1} is analyzed  as follows: The complexity of Algorithm~\ref{alg1} mainly depends on the calculation  of  ${\bm \pi}^G\left( {{{\bf{v}}^t}} \right)$ in step 3, the calculation  of picking the optimal vertex from the sequences  that maximizes the objective value  in   step 6, and the total number of iterations required to converge.  In the $t$th iteration,  the complexity of computing ${\bm \pi}^G\left( {{{\bf{v}}^t}} \right)$ by using Algorithm~\ref{alg2} is ${\cal O}\left( {{{\log }_2}\left( {{{{\lambda _{\max }} - {\lambda _{\min }}} \over \epsilon }} \right)\sqrt {M} } \right)$, where ${{{\log }_2}\left( {{{{\lambda _{\max }} - {\lambda _{\min }}} \over \epsilon}} \right)}$ is the number of iterations required for reaching convergence  by using the bisection method, and ${\sqrt M }$ is the complexity of solving linear optimization problem \eqref{P5} at each iteration in step 4 in Algorithm~\ref{alg2}.  For  each vertex,  the complexity for computing the objective function in  \eqref{P4} is ${\cal O}\left( M \right)$, thus the total  complexity of   step 6 is ${\cal O}\left( {\left( {tM - \left( {t - 1} \right)} \right)M} \right)$, where ${tM - \left( {t - 1} \right)}$ is the number of vertices. It was shown in \cite{zappone2017globally} that the total number of iterations, denoted as $T_c$, required for  convergence grows exponentially  with $M$, i.e., ${T_{{\rm{c}}}} = {\cal O}\left( {{2^M}} \right)$. Therefore, the total complexity of Algorithm~\ref{alg1} is ${\cal O}\left( {{T_{\rm{c}}}{{\log }_2}\left( {{{{\lambda _{\max }} - {\lambda _{\min }}} \over \epsilon}} \right)\sqrt M  + {{\left( {M + {T_{\rm{c}}}M - {T_{\rm{c}}} + 1} \right){T_{\rm{c}}}} \over 2}} \right)$.


We note that although a globally optimal solution for \eqref{P2} using the POA method is obtained, the computational complexity grows  exponentially  with the number of variables $M$. To address this issue, a lower-complexity SCA-based method  is discussed in the next section.
\section{Joint 3D trajectory and communication design optimization}
In this section, we investigate the joint 3D trajectory and communication design optimization for maximizing the system throughput using the low-complexity  SCA method. Problem  \eqref{P1} is a mixed integer and non-convex optimization problem due to the objective function \eqref{P1_const0},   constraints \eqref{P1_const1}, \eqref{P1_const2}, and \eqref{P1_const10}. We decompose  problem \eqref{P1} into three  sub-problems, and then optimize each sub-problem in an iterative way. Specifically, the three  sub-problems  are the communication scheduling optimization with fixed transmit power and 3D UAV trajectory; the 3D UAV trajectory  optimization with fixed transmit power and communication scheduling;  the transmit power optimization with fixed  communication scheduling and 3D UAV trajectory. First, we relax the integer communication scheduling constraints \eqref{P1_const1} and \eqref{P1_const2}  into continuous constraints  as
\begin{align}
&\sum\nolimits_{l = 1}^L {{x_l}\left[ n \right]}  \le 1,~0 \le {x_l}\left[ n \right] \le 1,\forall l,n,\label{SCA_relax1}\\
&\sum\nolimits_{k = 1}^K {{y_k}\left[ n \right]}  \le 1,~0 \le {y_k}\left[ n \right] \le 1, \forall k,n,\label{SCA_relax2}
\end{align}
\subsection{Communication scheduling optimization with fixed transmit power and trajectory}
For any given $Q$ and $P$, the communication scheduling sub-problem is given by
\begin{subequations} \label{SCA_P1}
\begin{align}
&\mathop {\max }\limits_{y_k[n],x_l[n]} {\beta _1}\sum\limits_{n = 1}^N {\sum\limits_{k = 1}^K {{y_k}\left[ n \right]} } {\log _2}\left( {1 + \frac{{{h_k}\left[ n \right]p_k^s\left[ n \right]}}{{f\left[ n \right]{p^u}\left[ n \right] + {\sigma ^2}}}} \right) +\notag\\
& {\beta _2}\sum\limits_{n = 1}^N {\sum\limits_{l = 1}^L x_l[n]{{{\log }_2}\left( {1 + \frac{{{g_l}[n]{p^u}\left[ n \right]}}{{\sum\limits_{k = 1}^K {{{\tilde h}_{k,l}}{y_k}\left[ n \right]p_k^s\left[ n \right] + {\sigma ^2}} }}} \right)} }\label{SCA_P1_const0}\\
&\quad{\rm s.t.}~\eqref{SCA_relax1},~\eqref{SCA_relax2}.
\end{align}
\end{subequations}
As can be seen, \eqref{SCA_P1_const0} is convex but not concave w.r.t to $y_k[n]$, which makes problem \eqref{SCA_P1} non-convex. To tackle it, we apply the SCA method \cite{boyd2004convex}. Specifically, for any  feasible point $y_k^r[n]$ in the $r$th iteration, we have
\begin{align}
R_l^{u}\left[ n \right] \ge & {\log _2}\left( {1 + \frac{{{g_l}[n]{p^u}\left[ n \right]}}{{\sum\nolimits_{k = 1}^K {{{\tilde h}_{k,l}}y_k^r\left[ n \right]p_k^s\left[ n \right] + {\sigma ^2}} }}} \right) -\notag\\ &\sum\limits_{k = 1}^K {A_k^l\left( {{y_k}\left[ n \right] - y_k^r\left[ n \right]} \right)}\overset{\triangle} {=} {\varphi ^{lb}}\left( {R_l^{u}\left[ n \right]} \right), \label{SCA_P1_0}
\end{align}
where $A_k^l = \frac{{{g_l}[n]{p^u}\left[ n \right]{{\tilde h}_{k,l}}p_k^s\left[ n \right]{{\log }_2}e}}{{\left( {\sum\limits_{k = 1}^K {{{\tilde h}_{k,l}}y_k^r\left[ n \right]p_k^s\left[ n \right] + {\sigma ^2}} } \right)\left( {\sum\limits_{k = 1}^K {{{\tilde h}_{k,l}}y_k^r\left[ n \right]p_k^s\left[ n \right] + {\sigma ^2}}  + {g_l[n]}{p^u}\left[ n \right]} \right)}}$. Obviously, ${\varphi ^{lb}}\left( {R_l^{u}\left[ n \right]} \right)$ is linear with $y_k[n]$, which is convex. Therefore, the value $y_k^{r+1}[n]$ in the $r{\rm +}1$th iteration can be achieved by solving the following convex problem:
\begin{subequations} \label{SCA_P1_1}
\begin{align}
&\mathop {\max }\limits_{y_k[n],x_l[n]} {\beta _1}\sum\limits_{n = 1}^N {\sum\limits_{k = 1}^K {{y_k}\left[ n \right]} } {\log _2}\left( {1 + \frac{{{h_k}\left[ n \right]p_k^s\left[ n \right]}}{{f\left[ n \right]{p^u}\left[ n \right] + {\sigma ^2}}}} \right) \notag\\
&\qquad\qquad +{\beta _2}\sum\limits_{n = 1}^N {\sum\limits_{l = 1}^L {x_l[n]{\varphi ^{lb}}\left( {R_l^{u}\left[ n \right]} \right)} }\label{SCA_P1_1_const0}\\
&{\rm s.t.}~\eqref{SCA_relax1},~\eqref{SCA_relax2}.\notag
\end{align}
\end{subequations}
By successively updating the $y_k^r[n]$, a locally optimal solution can be found.
\subsection{3D UAV trajectory  optimization with fixed transmit power and communication scheduling}
For any given $A$ and $P$, the 3D trajectory problem is given by
\begin{subequations} \label{SCA_P3}
\begin{align}
&\mathop {\max }\limits_{Q} {\beta _1}\sum\limits_{n = 1}^N {\sum\limits_{k = 1}^K {{y_k}\left[ n \right]} } {\log _2}\left( {1 + \frac{{{h_k}\left[ n \right]p_k^s\left[ n \right]}}{{f\left[ n \right]{p^u}\left[ n \right] + {\sigma ^2}}}} \right) + \notag\\
&{\beta _2}\sum\limits_{n = 1}^N {\sum\limits_{l = 1}^Lx_l[n] {{{\log }_2}\left( {1 + \frac{{{g_l}[n]{p^u}\left[ n \right]}}{{\sum\nolimits_{k = 1}^K {{{\tilde h}_{k,l}}{y_k}\left[ n \right]p_k^s\left[ n \right] + {\sigma ^2}} }}} \right)} } \label{SCA_P3_const0}\\
&\quad{\rm s.t.}~ \eqref{P1_const5}\text{-}\eqref{P1_const10}. 
\end{align}
\end{subequations}
Problem \eqref{SCA_P3} is non-convex due to the non-convex objective function \eqref{SCA_P3_const0} and non-convex constraint \eqref{P1_const10}. Let $\psi \left( {R_l^{u}\left[ n \right]} \right)$ be the first order Taylor expansion of $R_l^{u}[n]$ at the feasible point $Z_l^{u,r}[n]\overset{\triangle}{=}{{{\left\| {{\bf{q}}_u^r\left[ n \right] - {{\bf{w}}_{ul}}} \right\|}^2} + H_u^r{{\left[ n \right]}^2}}$ in  the $r$th iteration, which is given by
\begin{align}
&\psi \left( {R_l^{u}\left[ n \right]} \right) = {\log _2}\left( {1 + \frac{{{S_{1,l}}\left[ n \right]}}{{{{\left\| {{\bf{q}}_u^r\left[ n \right] - {{\bf{w}}_{ul}}} \right\|}^2} + H_u^r{{\left[ n \right]}^2}}}} \right) -{S_{2,l}}\left[ n \right] \notag\\
&\times\left( {{{\left\| {{{\bf{q}}_u}\left[ n \right] - {{\bf{w}}_{ul}}} \right\|}^2} + {H_u}{{\left[ n \right]}^2} - {{\left\| {{\bf{q}}_u^r\left[ n \right] - {{\bf{w}}_{ul}}} \right\|}^2} - H_u^r{{\left[ n \right]}^2}} \right),\label{SCA_P3_const1}
\end{align}
where ${S_{1,l}}\left[ n \right] = \frac{{{p^u}\left[ n \right]{\beta _0}}}{{\sum\nolimits_{k = 1}^K {{{\tilde h}_{k,l}}{y_k}\left[ n \right]p_k^s\left[ n \right] + {\sigma ^2}} }}$
and ${S_{2,l}}\left[ n \right] = \frac{{{S_{1,l}}\left[ n \right]}}{{Z_l^{u,r}\left[ n \right]\left( {Z_l^{u,r}\left[ n \right] + {S_{1,l}}\left[ n \right]} \right)}}$.
Equation \eqref{SCA_P3_const1} is concave w.r.t the UAV trajectory variable $Q$.
In addition,  $R_k^s\left[ n \right]$  in \eqref{SCA_P3_const0} can be rewritten as
\begin{align}
R_k^s\left[ n \right] = \hat R_k^s\left[ n \right] - \log \left( {\frac{{{\beta _0}{p^u}\left[ n \right]}}{{{{\left\| {{{\bf{w}}_u}\left[ n \right] - {{\bf{w}}_b}\left[ n \right]} \right\|}^2}}} + {\sigma ^2}} \right), \label{SCA_P3_const2}
\end{align}
where
\begin{align}
&\hat R_k^s\left[ n \right] = \log \left( {\frac{{{\beta _0}{p^u}\left[ n \right]}}{{{{\left\| {{{\bf{w}}_u}\left[ n \right] - {{\bf{w}}_b}\left[ n \right]} \right\|}^2}}} + } \right.\notag\\
&\qquad\qquad\left. {\frac{{{\beta _0}p_k^s\left[ n \right]}}{{{{\left\| {{{\bf{q}}_b}\left[ n \right] - {{\bf{w}}_{bk}}} \right\|}^2} + {H_b}{{\left[ n \right]}^2}}} + {\sigma ^2}} \right)\label{SCA_P3_const3}
\end{align}
By introducing the slack variables $\Upsilon \left[ n \right]$,  \eqref{SCA_P3_const2} can be recast as
\begin{align}
R_k^s\left[ n \right] = \hat R_k^s\left[ n \right] - \log \left( {\frac{{{\beta _0}{p^u}\left[ n \right]}}{{\Upsilon \left[ n \right]}} + {\sigma ^2}} \right), \label{SCA_P3_const4}
\end{align}
with the additional constraints
\begin{align}
0<\Upsilon \left[ n \right] \le {\left\| {{{\bf{w}}_u}\left[ n \right] - {{\bf{w}}_b}\left[ n \right]} \right\|^2},\forall n. \label{SCA_P3_const5}
\end{align}
We can see that the second term $\log \left( {\frac{{{\beta _0}{p^u}\left[ n \right]}}{{\Upsilon \left[ n \right]}} + {\sigma ^2}} \right)$ in \eqref{SCA_P3_const4} is convex w.r.t. $\Upsilon \left[ n \right]$. However, the new constraint \eqref{SCA_P3_const5} is non-convex. Let $\psi \left( {\Upsilon \left[ n \right]} \right)$ be the first order Taylor expansion of ${\left\| {{{\bf{w}}_u}\left[ n \right] - {{\bf{w}}_b}\left[ n \right]} \right\|^2}$ at the feasible point ${\bf{w}}_u^r\left[ n \right] = \left[ {{\bf{q}}_u^r\left[ n \right]{\kern 1pt} {\kern 1pt} {\kern 1pt} {\kern 1pt} H_u^r\left[ n \right]} \right],{\bf{w}}_b^r\left[ n \right] = \left[ {{\bf{q}}_b^r\left[ n \right]{\kern 1pt} {\kern 1pt} {\kern 1pt} {\kern 1pt} H_b^r\left[ n \right]} \right]$ in  the $r$th iteration. Then, we have
\begin{align}
&\psi \left( {\Upsilon \left[ n \right]} \right) = {\left\| {{\bf{w}}_u^r\left[ n \right] - {\bf{w}}_b^r\left[ n \right]} \right\|^2} + 2\left( {{\bf{w}}_u^r\left[ n \right] - {\bf{w}}_b^r\left[ n \right]} \right)\times \notag\\
&{\left( {{{\bf{w}}_u}\left[ n \right] - {\bf{w}}_u^r\left[ n \right]} \right)^T} -2\left( {{\bf{w}}_u^r\left[ n \right] - {\bf{w}}_b^r\left[ n \right]} \right){\left( {{{\bf{w}}_b}\left[ n \right] - {\bf{w}}_b^r\left[ n \right]} \right)^T}. \label{SCA_P3_const5.1}
\end{align}
The constraint \eqref{SCA_P3_const5}  can be reformulated as
\begin{align}
0 < \Upsilon \left[ n \right] \le \psi \left( {\Upsilon \left[ n \right]} \right),\forall n. \label{SCA_P3_const6}
\end{align}
Note that  the first  term $\hat R_k^s\left[ n \right]$ in \eqref{SCA_P3_const4} is also non-convex. To this end,  let $\psi \left( {\hat R_k^s\left[ n \right]} \right)$ be the first order Taylor expansion of $\hat R_k^s\left[ n \right]$ at any feasible points ${\left\| {{\bf{w}}_u^r\left[ n \right] - {\bf{w}}_b^r\left[ n \right]} \right\|^2}$ and ${{{\left\| {{\bf{q}}_b^r\left[ n \right] - {{\bf{w}}_{bk}}} \right\|}^2} + H_b^r{{\left[ n \right]}^2}}$ in  the $r$th iteration, which is given in \eqref{SCA_P3_const7}
\newcounter{mytempeqncnt1}
\begin{figure*}
	\normalsize
	\setcounter{mytempeqncnt1}{\value{equation}}
	\begin{align}
&\psi \left( {\hat R_k^s\left[ n \right]} \right) = \log \left( {\frac{{{\beta _0}{p^u}\left[ n \right]}}{{{{\left\| {{\bf{w}}_u^r\left[ n \right] - {\bf{w}}_b^r\left[ n \right]} \right\|}^2}}} + \frac{{{\beta _0}p_k^s\left[ n \right]}}{{{{\left\| {{\bf{q}}_b^r\left[ n \right] - {{\bf{w}}_{bk}}} \right\|}^2} + H_b^r{{\left[ n \right]}^2}}} + {\sigma ^2}} \right) - \notag\\
&{\Omega _{k,1}}\left[ n \right]\left( {{{\left\| {{{\bf{w}}_u}\left[ n \right] - {{\bf{w}}_b}\left[ n \right]} \right\|}^2} - {{\left\| {{\bf{w}}_u^r\left[ n \right] - {\bf{w}}_b^r\left[ n \right]} \right\|}^2}} \right) - {\Omega _{k,2}}\left[ n \right]\left( {{{\left\| {{{\bf{q}}_b}\left[ n \right] - {{\bf{w}}_{bk}}} \right\|}^2} + {H_b}{{\left[ n \right]}^2} - {{\left\| {{\bf{q}}_b^r\left[ n \right] - {{\bf{w}}_{bk}}} \right\|}^2} - H_b^r{{\left[ n \right]}^2}} \right)\label{SCA_P3_const7}
	%
	\end{align}
	\hrulefill 
	\vspace*{4pt} 
\end{figure*}
, where
\begin{align}
{\Omega _{k,1}}\left[ n \right] = \frac{{\frac{{{\beta _0}{p^u}\left[ n \right]}}{{{{\left\| {{\bf{w}}_u^r\left[ n \right] - {\bf{w}}_b^r\left[ n \right]} \right\|}^4}}}{{\log }_2}e}}{{\frac{{{\beta _0}{p^u}\left[ n \right]}}{{{{\left\| {{\bf{w}}_u^r\left[ n \right] - {\bf{w}}_b^r\left[ n \right]} \right\|}^2}}} + \frac{{{\beta _0}p_k^s\left[ n \right]}}{{{{\left\| {{\bf{q}}_b^r\left[ n \right] - {{\bf{w}}_{bk}}} \right\|}^2} + H_b^r{{\left[ n \right]}^2}}} + {\sigma ^2}}},
\end{align}
and
\begin{align}
{\Omega _{k,2}}\left[ n \right] = \frac{{\frac{{{\beta _0}p_k^s\left[ n \right]}}{{{{\left( {{{\left\| {{\bf{q}}_b^r\left[ n \right] - {{\bf{w}}_{bk}}} \right\|}^2} + H_b^r{{\left[ n \right]}^2}} \right)}^2}}}{{\log }_2}e}}{{\frac{{{\beta _0}{p^u}\left[ n \right]}}{{{{\left\| {{\bf{w}}_u^r\left[ n \right] - {\bf{w}}_b^r\left[ n \right]} \right\|}^2}}} + \frac{{{\beta _0}p_k^s\left[ n \right]}}{{{{\left\| {{\bf{q}}_b^r\left[ n \right] - {{\bf{w}}_{bk}}} \right\|}^2} + H_b^r{{\left[ n \right]}^2}}} + {\sigma ^2}}}.
\end{align}
In addition, the  constraint \eqref{P1_const10}  is non-convex. With \eqref{SCA_P3_const5.1},  constraint \eqref{P1_const10} can be replaced by
\begin{align}
\psi \left( {\Upsilon \left[ n \right]} \right) \ge d_{\min }^2,\forall n. \label{P1_const10_new}
\end{align}
As a result, with \eqref{SCA_P3_const1} and \eqref{SCA_P3_const7}, define the following optimization problem
\begin{subequations} \label{SCA_P3_1}
\begin{align}
&\mathop {\max }\limits_{Q,\Upsilon \left[ n \right]} {\beta _1}\sum\limits_{n = 1}^N {\sum\limits_{k = 1}^K {{y_k}\left[ n \right]} } \left( {\psi \left( {\hat R_k^{s}\left[ n \right]} \right) - \log \left( {\frac{{{\beta _0}{p^u}\left[ n \right]}}{{\Upsilon \left[ n \right]}} + {\sigma ^2}} \right)} \right)+\notag\\
&\qquad\qquad {\beta _2}\sum\limits_{n = 1}^N {\sum\limits_{l = 1}^Lx_l[n] {\psi \left( {R_l^{u}\left[ n \right]} \right)} } \\
&\quad{\rm s.t.}~ \eqref{P1_const5}\text{-}\eqref{P1_const9}, \eqref{SCA_P3_const6},\eqref{P1_const10_new}.
\end{align}
\end{subequations}
Problem \eqref{SCA_P3_1} can be efficiently solved by  standard methods due to its convexity. Then, a locally optimal solution to problem \eqref{SCA_P3} can be guaranteed by successively updating the 3D UAV trajectory obtained from  problem \eqref{SCA_P3_1}.
\subsection{Transmit power optimization with fixed  communication scheduling and 3D UAV trajectory}
For any given $A$ and $Q$, the transmit power optimization problem is simplified as
\begin{subequations} \label{SCA_P4}
\begin{align}
&\mathop {\max }\limits_{P} {\beta _1}\sum\limits_{n = 1}^N {\sum\limits_{k = 1}^K {{y_k}\left[ n \right]} } {\log _2}\left( {1 + \frac{{{h_k}\left[ n \right]p_k^s\left[ n \right]}}{{f\left[ n \right]{p^u}\left[ n \right] + {\sigma ^2}}}} \right) +\notag\\
&{\beta _2}\sum\limits_{n = 1}^N {\sum\limits_{l = 1}^Lx_l[n] {{{\log }_2}\left( {1 + \frac{{{g_l}[n]{p^u}\left[ n \right]}}{{\sum\nolimits_{k = 1}^K {{{\tilde h}_{k,l}}{y_k}\left[ n \right]p_k^s\left[ n \right] + {\sigma ^2}} }}} \right)} } \label{SCA_P4_const0}\\
&\quad {\rm s.t.}~ \eqref{P1_const3},\eqref{P1_const4}. 
\end{align}
\end{subequations}
The objective function \eqref{SCA_P4_const0} is non-convex. To tackle it, we again apply the SCA method. Specifically, we rewrite $R_k^s[n]$ as
\begin{align}
R_k^s\left[ n \right] = {\log _2}\left( {{h_k}\left[ n \right]p_k^s\left[ n \right] + f\left[ n \right]{p^u}\left[ n \right] + {\sigma ^2}} \right) - \tilde R_k^s\left[ n \right],\label{SCA_P4_const1}
\end{align}
where
\begin{align}
\tilde R_k^s\left[ n \right] = {\log _2}\left( {f\left[ n \right]{p^u}\left[ n \right] + {\sigma ^2}} \right).\label{SCA_P4_const2}
\end{align}
Obviously, \eqref{SCA_P4_const1} is a difference of convex (DC) function.  We replace the term $\tilde R_k^s\left[ n \right]$ by its   first order Taylor expansion  at  any given feasible point $p^{u,r}[n]$, denoted as $\psi \left( {\tilde R_k^s\left[ n \right]} \right)$, and  given by
\begin{align}
\psi \left( {\tilde R_k^s\left[ n \right]} \right) = &{\log _2}\left( {f\left[ n \right]{p^{u,r}}\left[ n \right] + {\sigma ^2}} \right) +\notag\\
&\frac{{f\left[ n \right]{{\log }_2}e}}{{f\left[ n \right]{p^{u,r}}\left[ n \right] + {\sigma ^2}}}\left( {{p^u}\left[ n \right] - {p^{u,r}}\left[ n \right]} \right).  \label{SCA_P4_const3}
\end{align}
Next, we tackle the non-convexity of $R_l^{u}\left[ n \right]$ in \eqref{SCA_P4_const0} by  rewriting $R_l^{u}\left[ n \right]$ as
\begin{align}
R_l^{u}\left[ n \right] = &{\log _2}\left( {{g_l}[n]{p^u}\left[ n \right] + \sum\limits_{k = 1}^K {{{\tilde h}_{k,l}}{y_k}\left[ n \right]p_k^s\left[ n \right] + {\sigma ^2}} } \right)\notag\\
&-\tilde R_l^{u}\left[ n \right],\label{SCA_P4_const4}
\end{align}
where
\begin{align}
\tilde R_l^{u}\left[ n \right] = {\log _2}\left( {\sum\limits_{k = 1}^K {{{\tilde h}_{k,l}}{y_k}\left[ n \right]p_k^s\left[ n \right] + {\sigma ^2}} } \right).\label{SCA_P4_const5}
\end{align}
Interestingly, \eqref{SCA_P4_const4} is also a difference of convex (DC) functions. By taking the  same steps as in \eqref{SCA_P4_const1}, an  upper bound for  $\tilde R_l^{u}\left[ n \right]$ at any feasible point ${p_k^{s,r}\left[ n \right]}$ is given by
\begin{align}
&\psi \left( {\tilde R_l^{u}\left[ n \right]} \right) = {\log _2}\left( {\sum\nolimits_{k = 1}^K {{{\tilde h}_{k,l}}{y_k}\left[ n \right]p_k^{s,r}\left[ n \right] + {\sigma ^2}} } \right) + \notag\\
&\sum\limits_{k = 1}^K {\frac{{{{\tilde h}_{k,l}}{y_k}\left[ n \right]}}{{\sum\nolimits_{k = 1}^K {{{\tilde h}_{k,l}}{y_k}\left[ n \right]p_k^{s,r}\left[ n \right] + {\sigma ^2}} }}\left( {p_k^s\left[ n \right] - p_k^{s,r}\left[ n \right]} \right)}. \label{SCA_P4_const6}
\end{align}
Consequently, with \eqref{SCA_P4_const3} and \eqref{SCA_P4_const6}, we define the optimization problem in \eqref{SCA_P4_1}.
\newcounter{mytempeqncnt2}
\begin{figure*}
	\normalsize
	\setcounter{mytempeqncnt2}{\value{equation}}
\begin{subequations} \label{SCA_P4_1}
	\begin{align}
	&\mathop {\max }\limits_P {\beta _1}\sum\limits_{n = 1}^N {\sum\limits_{k = 1}^K {{y_k}\left[ n \right]} } \Bigg( {{{\log }_2}\left( {{h_k}\left[ n \right]p_k^s\left[ n \right] + f\left[ n \right]{p^u}\left[ n \right] + {\sigma ^2}} \right)}-{\psi \left( {\tilde R_k^s\left[ n \right]} \right)} \Bigg)   \notag \\
	&\qquad + {\beta _2}\sum\limits_{n = 1}^N {\sum\limits_{l = 1}^L {{x_l}\left[ n \right]} } \Bigg( {{{\log }_2}\Big( {{g_l}[n]{p^u}\left[ n \right] + }}{\sum\limits_{k = 1}^K {{{\tilde h}_{k,l}}{y_k}\left[ n \right]p_k^s\left[ n \right] + {\sigma ^2}} } \Big) -  {\psi \left( {\tilde R_l^{u}\left[ n \right]} \right)} \Bigg) \label{SCA_P4_1_const0}\\
	&\quad {\rm s.t.}~ \eqref{P1_const3},\eqref{P1_const4}. 
	\end{align}
\end{subequations}
	\hrulefill 
	\vspace*{4pt} 
\end{figure*}
It can be verified that problem \eqref{SCA_P4_1} is a convex optimization problem, which can be readily solved. Then, a locally optimal solution to problem \eqref{SCA_P4} can be guaranteed by successively updating the transmit power  obtained from  problem \eqref{SCA_P4_1}.
\begin{algorithm}[!t]
\caption{BCD for problem \eqref{P1}.}\label{alg3}
\begin{algorithmic}[1]
\STATE  \textbf{Initialize} ${P^r}$, ${Q^r}$, and set $r=0$ as well as tolerance $\epsilon=10^{-2}$.
\STATE  \textbf{repeat}
\STATE  \quad Solve  \eqref{SCA_P1_1} for given ${P^r}$ and ${Q^r}$, and denoted the \\
\quad optimal solution as ${A^{r + 1}}$.
\STATE  \quad Solve  \eqref{SCA_P3_1} for given ${A^{r+1}}$ and ${P^r}$, and denoted the \\
\quad optimal solution as ${Q^{r + 1}}$.
\STATE  \quad Solve  \eqref{SCA_P4_1} for given ${A^{r+1}}$ and  ${Q^{r + 1}}$, and denoted the \\
\quad optimal solution as ${P^{r + 1}}$.
\STATE  \quad $r=r+1$.
\STATE \textbf{until}  the fractional increase of the objective value of \eqref{P1} is smaller than $\epsilon$.
\end{algorithmic}
\end{algorithm}
\subsection{Overall algorithm}
Based on the solutions to the  three sub-problems above,  we  alternately optimize the three sub-problems based on the  block coordinate descent (BCD) method \cite{wang2018joint}, \cite{wu2018Joint},\cite{spectrum2018wang}.  The details of the BCD are summarized in Algorithm~\ref{alg3}.

The convergence of Algorithm~\ref{alg3} is proved as follows: To facilitate the design, we define ${A^{r + 1}} = \left\{ {x_l^r\left[ n \right],y_k^r\left[ n \right],\forall l,k,n} \right\}$, ${P^r} = \left\{ {{p^{u,r}}\left[ n \right],p_k^{s,r}\left[ n \right],\forall k,n} \right\}$, and $Q^r = \left\{ {{\bf{w}}_u^r\left[ n \right],{\bf{w}}_b^r\left[ n \right],\forall n} \right\}$ in the $r$th iteration. Let  $R\left( {{A^r},{Q^r},{P^r}} \right)$, $R_u^{lb}\left( {{A^r},{Q^r},{P^r}} \right)$, $R_q^{lb}\left( {{A^r},{Q^r},{P^r}} \right)$, and $R_p^{lb}\left( {{A^r},{Q^r},{P^r}} \right)$ be the objective value to the relaxed problem \eqref{P1}, \eqref{SCA_P1_1}, \eqref{SCA_P3_1}, and \eqref{SCA_P4_1} in the $(r+1)$th iteration, respectively. In the $r$th iteration, in step 3 of Algorithm~\ref{alg3}, we  have
\begin{align}
R\left( {{A^r},{Q^r},{P^r}} \right)&\overset{(a)}{=}R_u^{lb}\left( {{A^r},{Q^r},{P^r}} \right) \overset{(b)} {\le} R_u^{lb}\left( {{A^{r + 1}},{Q^r},{P^r}} \right) \notag\\
&\overset{(c)}{\le} R\left( {{A^{r + 1}},{Q^r},{P^r}} \right), \label{convergence1}
\end{align}
where (a) holds since the first-order Taylor expansion in \eqref{SCA_P1_0} is tight at the given local point $A^r$, which indicates that problem \eqref{SCA_P1_1} at $A^r$ has the same objective value as that of problem \eqref{SCA_P1}; (b) holds since in step 3  with the given $Q^r$ and $P^r$, problem \eqref{SCA_P1_1} is solved optimally with solution $A^{r+1}$; and (c) holds due to that the objective value of \eqref{SCA_P1_1} is served as a  lower bound  to   that of \eqref{SCA_P1}. The inequality \eqref{convergence1} shows that the objective value of \eqref{SCA_P1} is non-decreasing after each iteration. Similar to step 4 and step 5, we respectively have
\begin{align}
R\left( {{A^{r + 1}},{Q^r},{P^r}} \right)& = R_q^{lb}\left( {{A^{r + 1}},{Q^r},{P^r}} \right) \le R_q^{lb}\left( {{A^{r + 1}},{Q^{r + 1}},{P^r}} \right) \notag\\
&\le R\left( {{A^{r + 1}},{Q^{r + 1}},{P^r}} \right), \label{convergence2}
\end{align}
and
\begin{align}
&R\left( {{A^{r + 1}},{Q^{r + 1}},{P^r}} \right) = R_p^{lb}\left( {{A^{r + 1}},{Q^{r + 1}},{P^r}} \right)\notag\\
 &\le R_p^{lb}\left( {{A^{r + 1}},{Q^{r + 1}},{P^{r + 1}}} \right) \le R\left( {{A^{r + 1}},{Q^{r + 1}},{P^{r + 1}}} \right). \label{convergence3}
\end{align}
Based on \eqref{convergence1}\text{-}\eqref{convergence3}, we obtain the following inequality
\begin{align}
R\left({{A^r},{Q^r},{P^r}} \right) \le R\left( {{A^{r + 1}},{Q^{r + 1}},{P^{r + 1}}} \right),\label{convergence4}
\end{align}
which shows that the objective value of the relaxed problem \eqref{P1} is non-decreasing after each  iteration. In addition, the maximum objective value of problem \eqref{P1} is upper bounded  by  a finite value due to the limited flying time and UAV-AP/SN transmit power budget in practice. Therefore, Algorithm~\ref{alg3} is guaranteed to converge to a locally optimal solution.  Note that Algorithm~\ref{alg3}  solves  the relaxed problem \eqref{P1}, where the binary communication scheduling is relaxed to the continuous variables between 0 and 1. To reconstruct the binary communication scheduling, we directly apply  the rounding operation adopted  in  \cite{hua2019energy}, \cite{hua2019fulluav}.

Next, we analyze the complexity of Algorithm~\ref{alg3}. In step 3 of Algorithm~\ref{alg3},  sub-problem \eqref{SCA_P1_1} is a linear optimization problem, which can be solved by the  interior point method  with computational  complexity ${\cal O}\left( {L_1\sqrt {KN+LN} } \right)$ \cite{gondzio1996computational}, where  $KN+LN$ denotes the total number of  variables, and  $L_1$ denotes the number of iterations required to update the communication scheduling. In step 4, since sub-problem \eqref{SCA_P3_1} involves the logarithmic form, the  complexity for solving \eqref{SCA_P3_1} by  using the interior point method is ${\cal O}\left( {{L_2}{{\left( {7N} \right)}^{3.5}}} \right)$ \cite{zhang2019securing}, where $7N$ represents  the total number of  variables, and $L_2$ denotes the number of iterations required to update the UAV trajectory. Similarly, sub-problem \eqref{SCA_P4_1} also  involves the logarithmic form,  the complexity is ${\cal O}\left(L_3\left(KN+N\right)^{3.5}\right)$, where $L_3$ represents the number of iterations required to update the transmit power, and $KN+N$ stands for the number of variables. Therefore, the overall complexity of Algorithm~\ref{alg3} is ${\cal O}\left( {{L_4}\left( {{L_1}\left( {\sqrt {KN + LN} } \right){\rm{ + }}{L_2}{{\left( {7N} \right)}^{3.5}}{\rm{ + }}{L_3}\left( {KN{\rm{ + }}N} \right)}^{3.5} \right)} \right)$ with $L_4$ being the number of iterations required by Algorithm~\ref{alg3} to converge.

\section{NUMERICAL RESULTS}
In this section,  numerical examples are provided to validate the effectiveness of the proposed algorithms. Unless otherwise specified, the simulation parameters are set as follows. We assume that the system bandwidth is $B{\rm =}1\rm{ MHz}$ with noise power $\sigma^2{\rm =-110}{\rm dBm}$ \cite{zeng2017energy}.  The G2G channel gain is  ${\beta _0}{\rm =-60}\rm dBm$ with path loss exponent $\alpha=3$ \cite{lyu2018uav}. The UAV altitude constraints are $H_{\rm min}{\rm =}100\rm{m}$ and $H_{\rm max}{\rm =}600\rm{m}$. The maximum horizontal and vertical  UAV speed are set to $V_{xy }{\rm =}50{\rm{m/s}}$ and $V_{z }{\rm =}30{\rm{m/s}}$, respectively. The minimum safety distance between two UAVs is  $d_{\rm min}{\rm =}10\rm m$. The maximum UAV-AP and SN transmit power is set as  $p_{\rm max}^s{\rm =}0.1{\rm W}$ and $p_{\rm max}^u{\rm =}0.1{\rm W}$, respectively. In addition, the  duration of each time slot is set as $\delta{\rm =}0.5{\rm s}$, and  the  penalty factor is set as $M=1\times10^5$.
\begin{figure}[!t]
\centering
\begin{minipage}[t]{0.45\textwidth}
\centering
\includegraphics[width=2.8in]{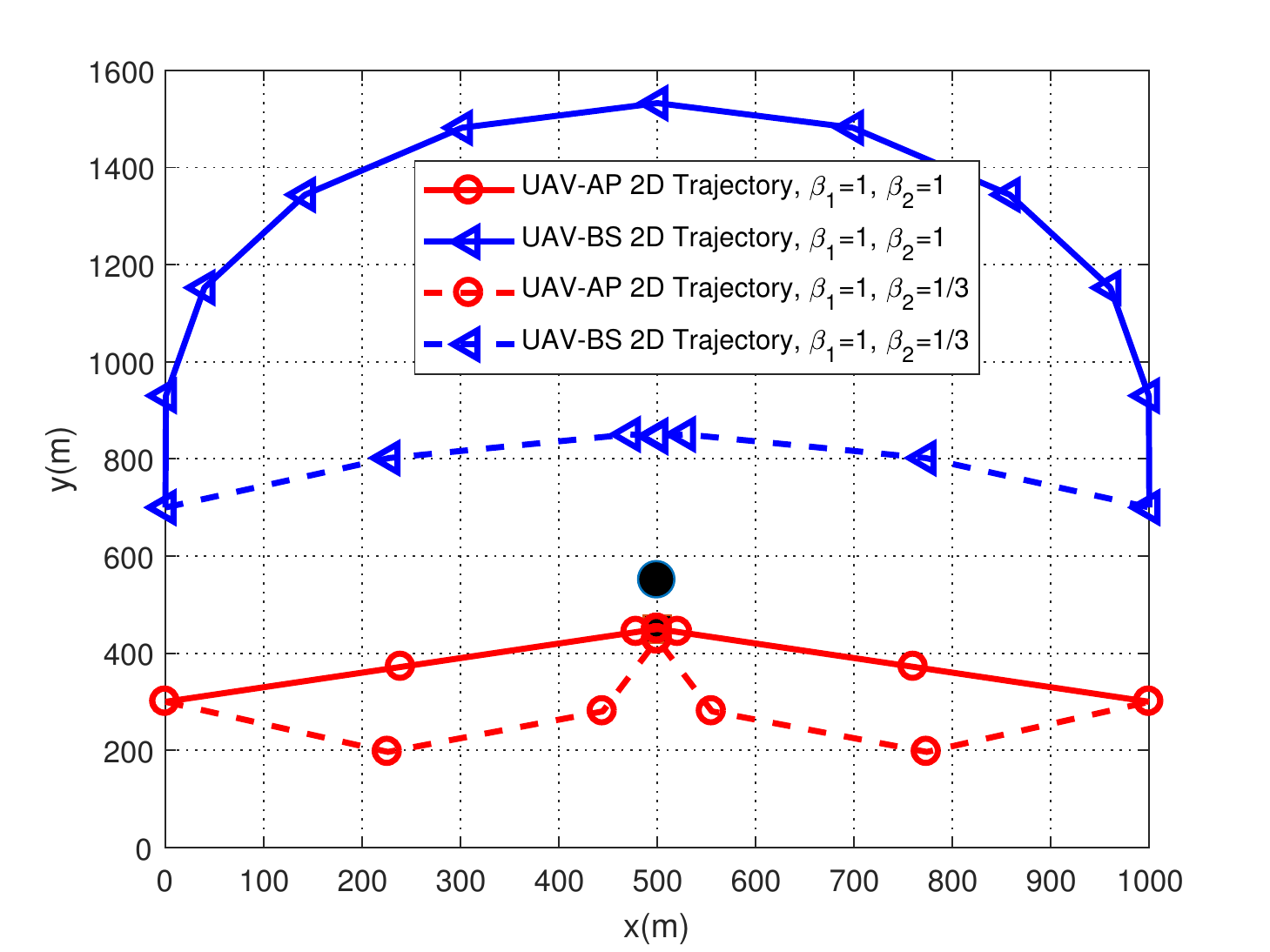}
\caption{Optimized horizontal UAV trajectories.}\label{SOSOfig1}
\end{minipage}
\begin{minipage}[t]{0.45\textwidth}
\centering
\includegraphics[width=2.8in]{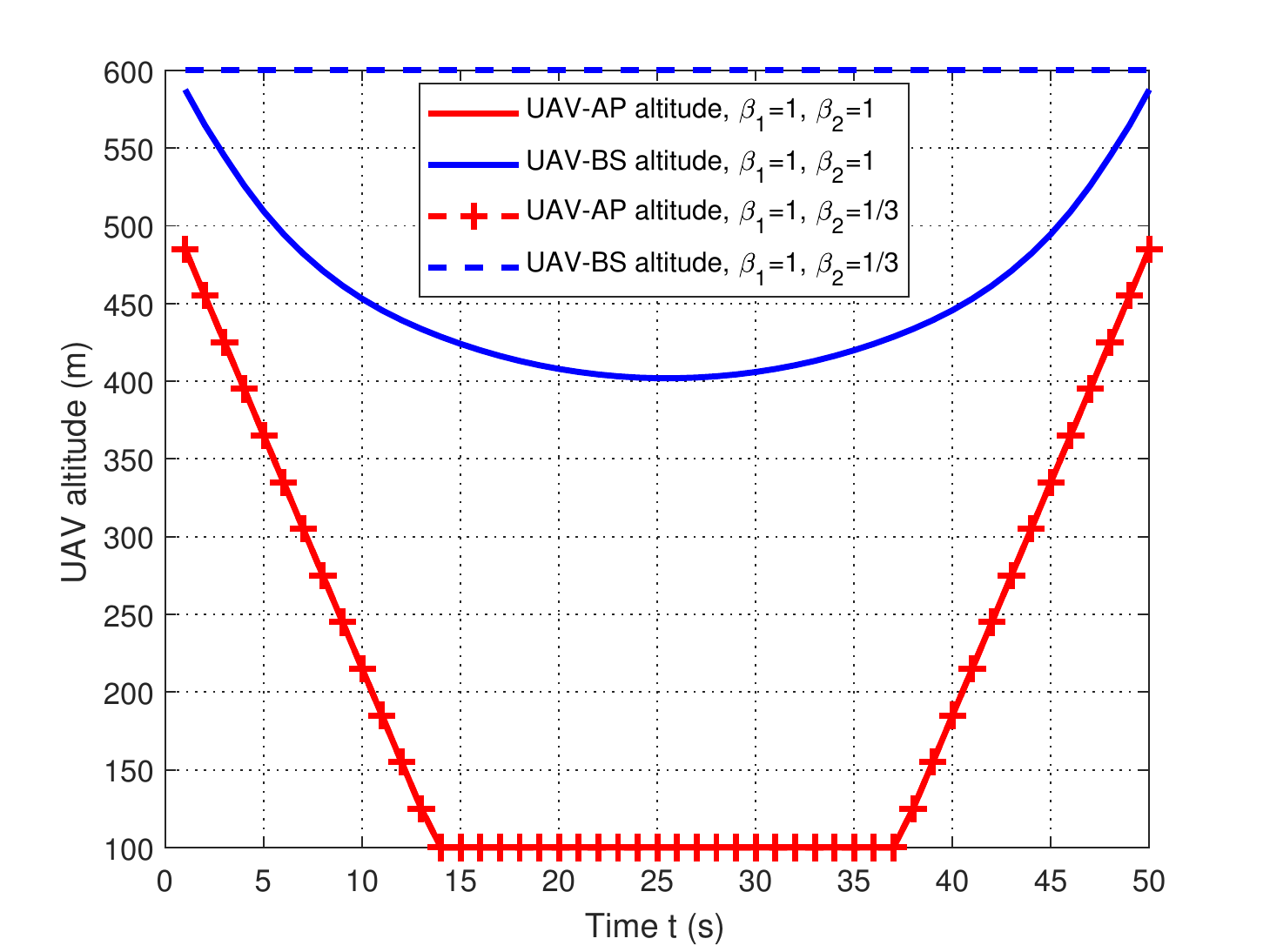}
\caption{Optimized UAV altitudes. }\label{SOSOfig2}
\end{minipage}
\end{figure}
\subsection{Single SN and single AP case}
We first consider a simple case where the UAV-BS collects
 data from one SN, and the UAV-AP transmits  its data to one AP.
The initial locations of UAVs and AP/SN are set as follows: ${\bf q}_{I_u}$=[0 300$\rm m$], ${\bf q}_{I_b}$=[0 700$\rm m$], ${\bf q}_{F_u}$=[1000$\rm m$ $300\rm m$], ${\bf q}_{F_b}$=[1000$\rm m$ 700$\rm m$],   ${\bf w}_{b1}$=[500$\rm m$ 550$\rm m$], ${\bf w}_{u1}$=[500$\rm m$ 450$\rm m$], $H_b$=600$\rm m$, and $H_u$=500$\rm m$.

In Fig.~\ref{SOSOfig1} and Fig.~\ref{SOSOfig2}, we plot the UAV-BS and UAV-AP 3D trajectories obtained by the  SCA method for different weighting  factors $\beta_2{\rm=}1$ and $\beta_2{\rm=}1/3$ when  $T=50s$. In Fig.~\ref{SOSOfig1}, it is observed that both UAVs remain separated from  each other to alleviate the interference received by  the UAV-BS from the UAV-AP. In addition, as $\beta_2$ becomes smaller, the UAV-BS prefers moving closer to the SN, since the UAV-BS  system throughput can be  significantly improved by establishing a better channel between the UAV-BS and the SN. In addition, the UAV-AP tends to move far from the  UAV-BS to reduce the interference imposed on the UAV-BS-based network. Finally, we can observe from Fig.~\ref{SOSOfig2} that   under $\beta_1=1$ and $\beta_2=1$, both  UAVs  descend to reduce  the  path loss and improve the system throughput.
\begin{figure}[!t]
\setlength{\abovecaptionskip}{0pt}
\setlength{\belowcaptionskip}{10pt}
\centering
\begin{minipage}[t]{0.45\textwidth}
\centering
\includegraphics[width=2.8in]{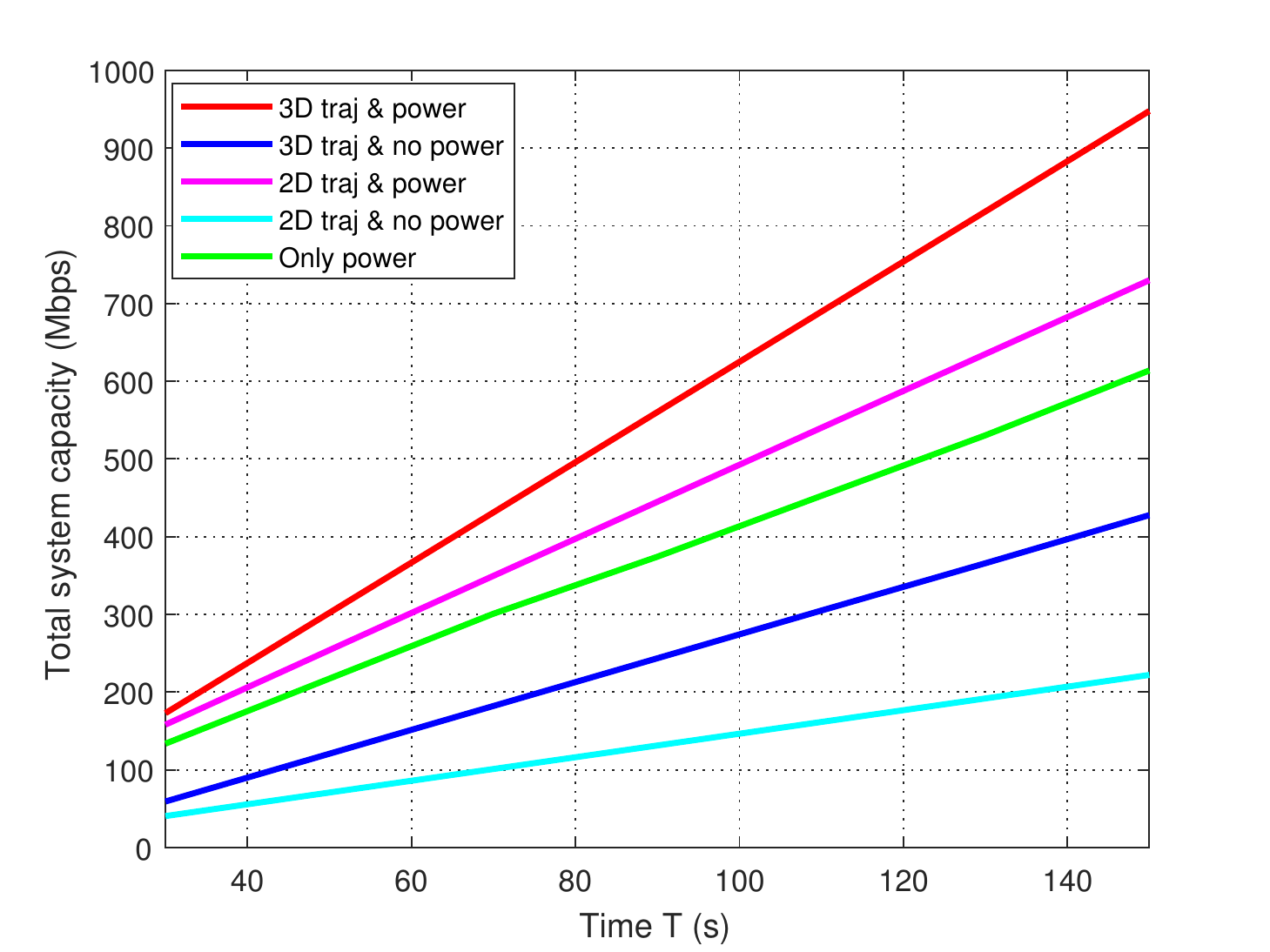}
\caption{System throughput versus period time $T$.}\label{SOSOfig4}
\end{minipage}
\begin{minipage}[t]{0.45\textwidth}
\centering
\includegraphics[width=2.8in]{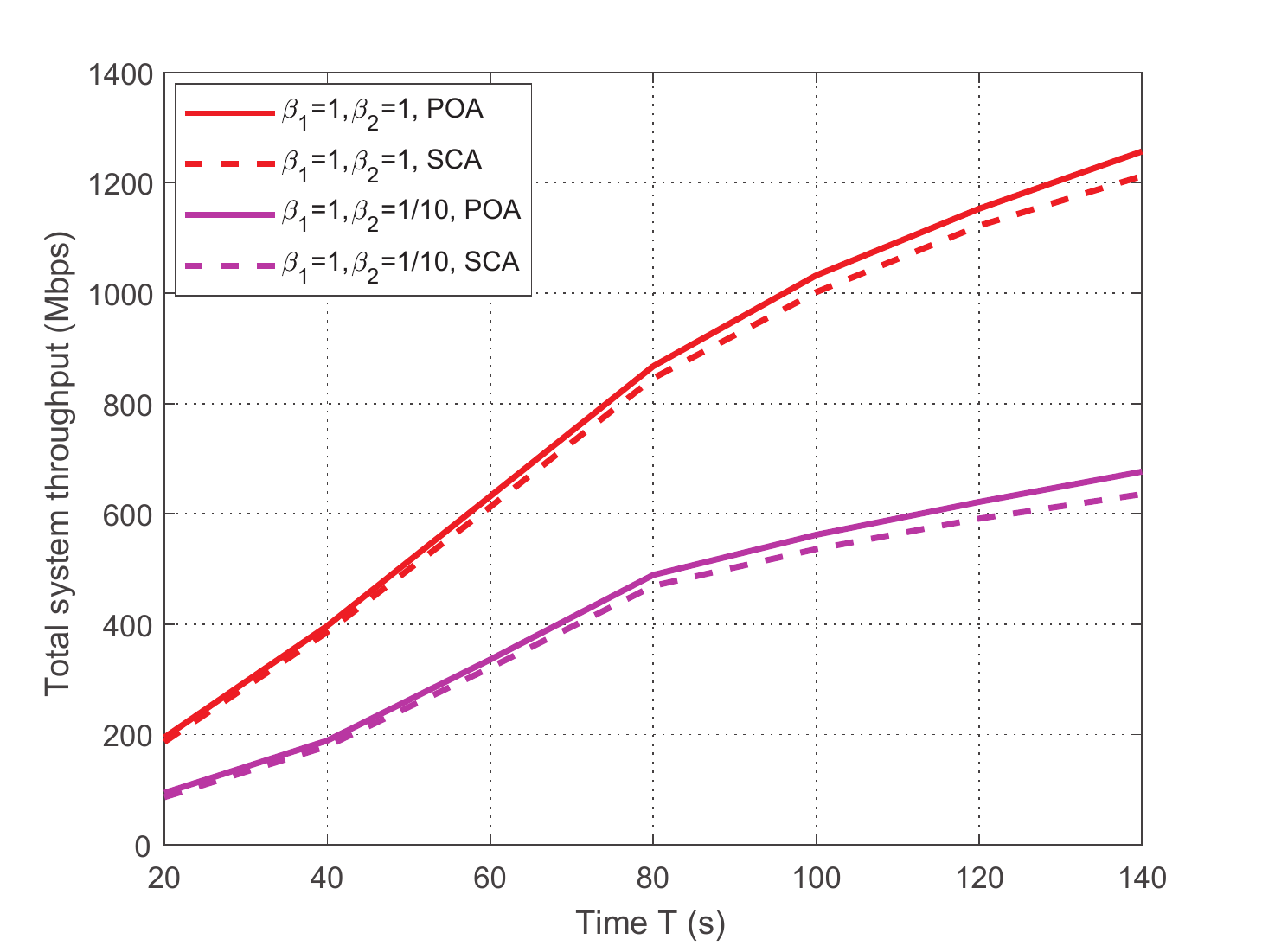}
\caption{Total system throughput versus period $T$ for different weighting  factors  using POA and SCA methods.}\label{Communicationdesignfig1}
\end{minipage}
\end{figure}

In Fig.~\ref{SOSOfig4}, we investigate the total system throughput versus period $T$ under $\beta_1=1$ and $\beta_2=1/3$ for different benchmarks to show the superiority of our proposed scheme. The definitions of the abbreviations of the benchmarks are given as below: 1) ``3D traj \& power'': This is our proposed scheme that jointly optimizes the 3D UAV trajectory and communication design; 2) ``3D traj \& no power'': The  3D UAV trajectory and communication scheduling are jointly optimized, but the transmit power is fixed at maximum power $p_{\rm max}^s=p_{\rm max}^u=0.1 \rm{W}$; 3) ``2D traj \& power'': The UAV altitude is fixed, the horizontal UAV trajectory and communication design are jointly optimized; 4) ``2D traj \& no power'': The  2D UAV trajectory and communication scheduling are jointly optimized, but  the transmit power of the UAV/SN and altitude of the UAV are fixed ($p_{\rm max}^s=p_{\rm max}^u=0.1 \rm{W}$); 5) ``Only power'': The UAV  horizontal trajectory and altitude are predetermined, the horizontal trajectory for the UAV-AP/UAV-BS is a straight line  from its initial location to its final location with constant  speed). However, the communication design, including communication scheduling and transmit power, is optimized.
%
%
%
%
%
First, we observe that our proposed scheme is superior to the other benchmarks and achieves significant throughput gains, especially when the period becomes larger. Second, the system throughput can be improved by controlling the UAV altitude. For instance, for period $T=130s$, the system throughput for the proposed scheme  is  818$\rm Mbps$, and for the ``2D trajectory \& power'' method is 634$\rm Mbps$, which provides  a nearly  23\% increase. In addition, the system throughput can be significantly improved by controlling the transmit power. For example, for period $T=130s$,  the system throughput for the ``3D trajectory \& no power''  method  is 365$\rm Mbps$, and for the ``2D trajectory \& no power''  method  is 191$\rm Mbps$, which correspond to a  55\% and  76\% increase in  the system throughput, respectively. Finally, the UAV trajectory design also has significantly impacts on the system performance.  For example, for period $T=130s$, the system throughput for  the ``only power'' method  is  530$\rm Mbps$, which results in  a 35\% increase in  the system throughput compared with our proposed method.

\subsection{Multiple SNs and multiple APs case}
In this section, we consider a more practical case where the UAV-BS and UAV-AP simultaneously serve multiple SNs and APs. The communication design, including power control and communication scheduling, and UAV trajectory are optimized.  We consider $4$ SNs and $4$ APs, which are  respectively located at   ${\bf w}_{b1}$=[-1000$\rm m$ 0], ${\bf w}_{b2}$=[-100$\rm m$ 700$\rm m$], ${\bf w}_{b3}$=[0~0], ${\bf w}_{b4}$=[-500$\rm m$ -500$\rm m$],  ${\bf w}_{u1}$=[1000$\rm m$ 0], ${\bf w}_{u2}$=[0 700$\rm m$], ${\bf w}_{u3}$=[100$\rm m$ 0], ${\bf w}_{u4}$=[700$\rm m$ -400$\rm m$].

In Fig.~\ref{Communicationdesignfig1}, we compare the total sum system throughput achieved by the POA and SCA method, versus period $T$, for different weighting  factors. The initial trajectories  for the UAV-BS and UAV-AP are   circles with given radii and centers. Specifically,   for any given period $T$ and   maximum UAV horizontal speed $V_{xy}$, the circle radius is first calculated by ${r_c} = \frac{{{V_{xy}}T}}{{2\pi }}, c \in\{b,u\}$. Second, for  any given location of  ${\bf w}_{ci}, c \in\{b,u\}, i \in\{{\cal K},{\cal L}\}$, the geometric center of the SNs and APs are ${{\bf ge}_b} = \frac{{\sum\limits_{i = 1}^K {{{\bf{w}}_{bi}}} }}{K}=[x_b~y_b]$ and ${{\bf ge}_u} = \frac{{\sum\limits_{i = 1}^L {{{\bf{w}}_{ui}}} }}{L}=[x_u~y_u]$, respectively.
Here, we consider two different weighting factors:  $\beta_1=1,\beta_2=1$ and  $\beta_1=1,\beta_2=1/10$.  As can be seen, when period $T$ is small, namely $T\le80s$, the system throughput obtained by the POA-based and SCA-based method is nearly the same both for the two different weighting factors. Even as  $T$ becomes larger, the throughput gap between the two algorithms still remains quite small.   In addition, for  $T=80s$ under $\beta_1=1,\beta_2=1/10$, the running time for the SCA-based method is about 3.7 minutes, while for the  POA-based method is nearly 27 hours. This indicates that   the  SCA based method can achieve  nearly the same optimal performance of  the POA-based method while with  much lower computational complexity.

\begin{figure}[!t]
\centering
\begin{minipage}[t]{0.45\textwidth}
\centering
\includegraphics[width=2.8in]{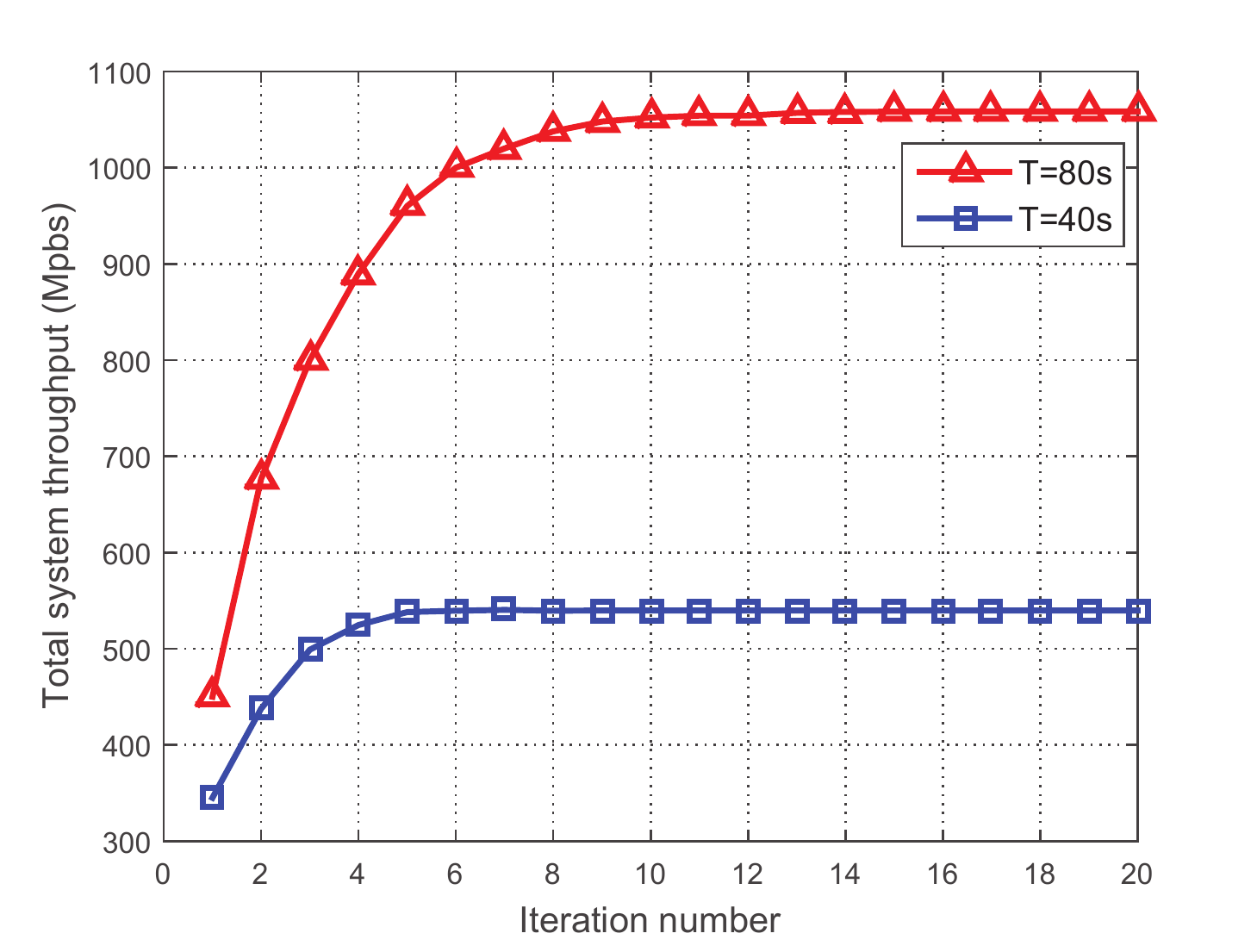}
\caption{Convergence behavior of Algorithm~\ref{alg3}.}\label{convergence}
\end{minipage}
\begin{minipage}[t]{0.45\textwidth}
\centering
\includegraphics[width=2.8in]{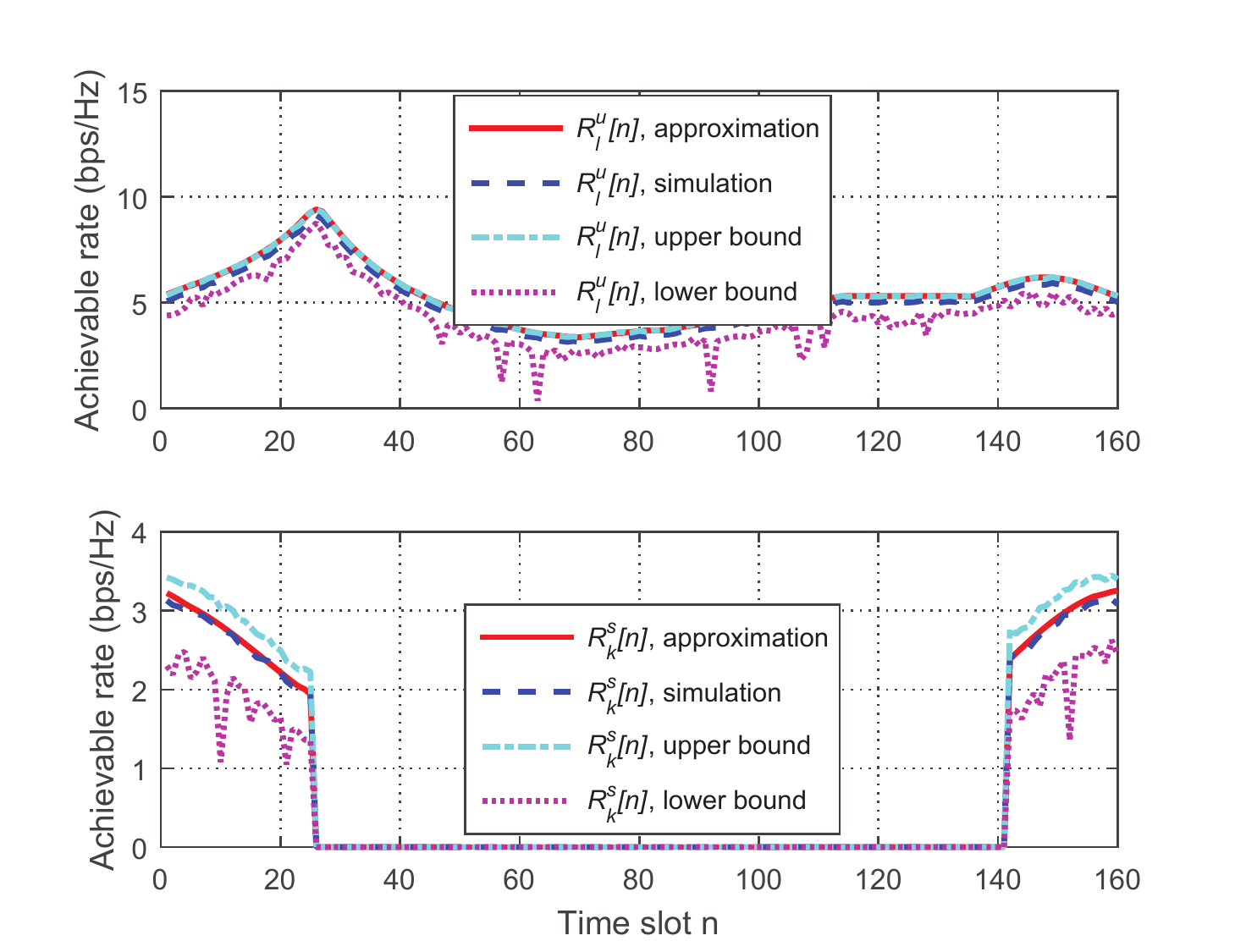}
\caption{Average communication throughput based on approximation versus numerical simulations. }\label{approximateresults}
\end{minipage}
\end{figure}

In Fig.~\ref{convergence}, we show the convergence behavior of Algorithm~\ref{alg3} for  different periods $T$, namely   $T=80s$ and  $T=40s$,  under  $\beta_1=\beta_2=1$. It is observed that the  system  throughput  obtained by the different periods $T$ all  increases quickly with the number of iterations. For a small  period $T=40s$, the proposed algorithm converges in about $6$ iterations, while for a large period $T=80s$, only $10$ iterations is required for achieving  convergence,  which demonstrates the efficiency of Algorithm~\ref{alg3}.

In order to evaluate the accuracy of the approximation of the expected  throughput both for uplink and downlink, i.e., $R_k^s[n]$ and $R_u^l[n]$, $\forall k,l,n$, developed  in \eqref{ergodicR_ks} and \eqref{ergodicR_ul}, the average throughput based on $R_k^s\left[ n \right] = {\mathbb E}\left\{ {\bar R_k^s\left[ n \right]} \right\}$ and $R_l^u\left[ n \right] = {\mathbb E}\left\{ {\bar R_l^u\left[ n \right]} \right\}$ obtained via numerical simulations is compared. Fig.~\ref{approximateresults} shows the results of approximation and numerical simulations under Rician factors $K_a=K_s=K_u=3\rm dB$, $T=80s$, and  $\beta_1=\beta_2=1$. Since there are multiple SNs and APs, we pick one SN and one AP, and compare  the results of  $R_1^s[n]$ and $R_1^u[n]$ without loss of generality. In addition, for the numerical simulation of the average throughput, the UAV trajectory and UAV-AP/SN transmit power are set as that obtained via Algorithm~\ref{alg3}, and the average throughput is taken over $10^4$ random channel generations at each time slot. It is observed that the approximation results match well with the  simulation results for both SN's throughput and AP's throughput at any time slot. In addition, the upper bound and lower bound of the average throughput obtained via numerical simulations are also plotted. One can see that the obtained approximation results indeed lie in the interval between   them, which are consistent with   \eqref{appendix1_const2} and \eqref{appendix1_const4} in Appendix~\ref{appendix1}. It should be noted that the curves of  lower bound results are not smooth, and  fluctuate drastically in some time slots. This is because in some time slots, the numerator in the logarithm form  approaches nearly zero (See $X$ in the left hand side of    \eqref{appendix1_const2}  and \eqref{appendix1_const4}).

\begin{figure*}[!t]
\centering
\subfigure[$\beta_1=\beta_2=1$.]{
\begin{minipage}[t]{0.45\linewidth}
\centering
\includegraphics[width=2.8in]{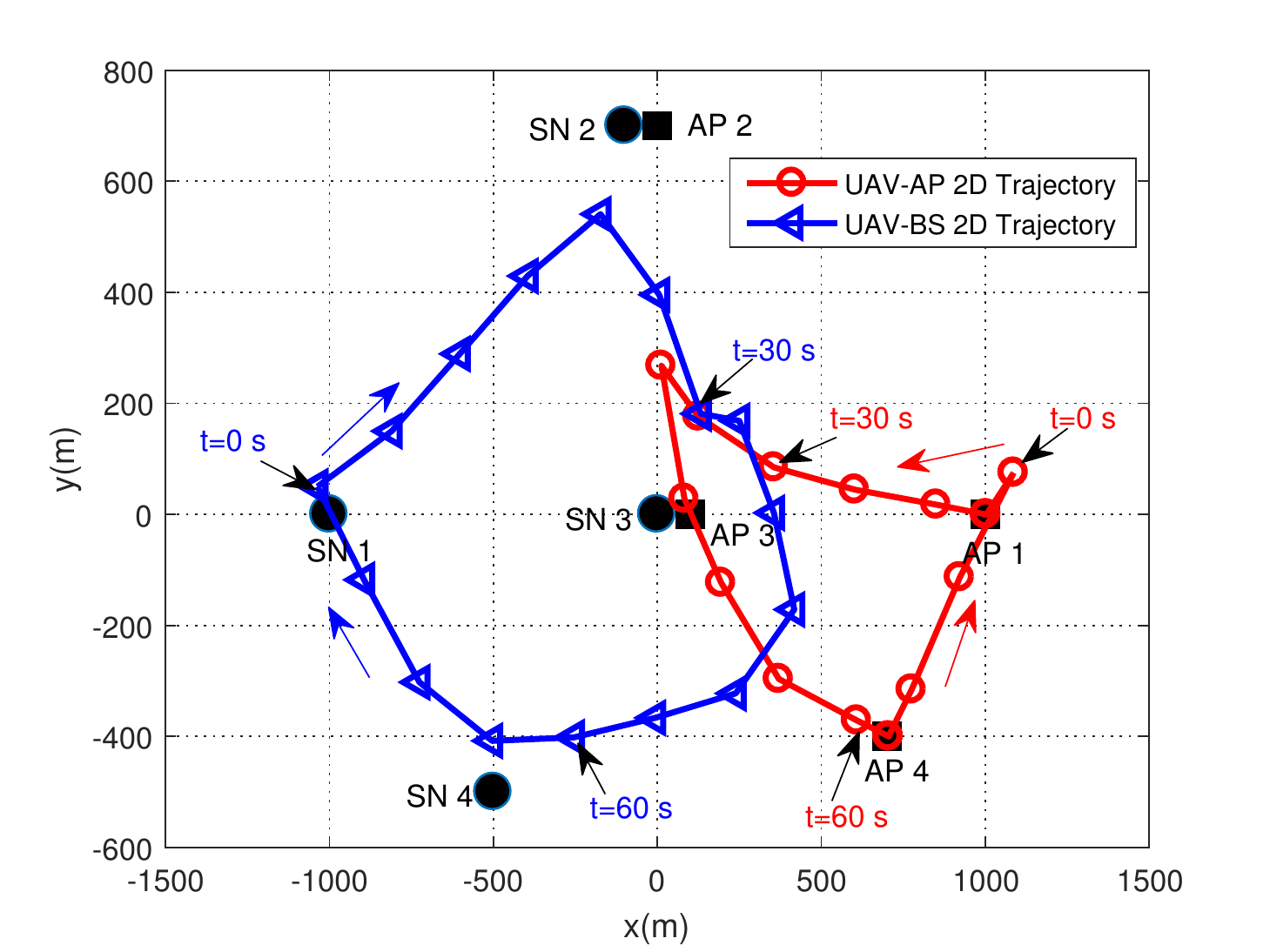}
\end{minipage}%
}%
\quad               
\subfigure[$\beta_1=1, \beta_2=1/10$.]{
\begin{minipage}[t]{0.45\linewidth}
\centering
\includegraphics[width=2.8in]{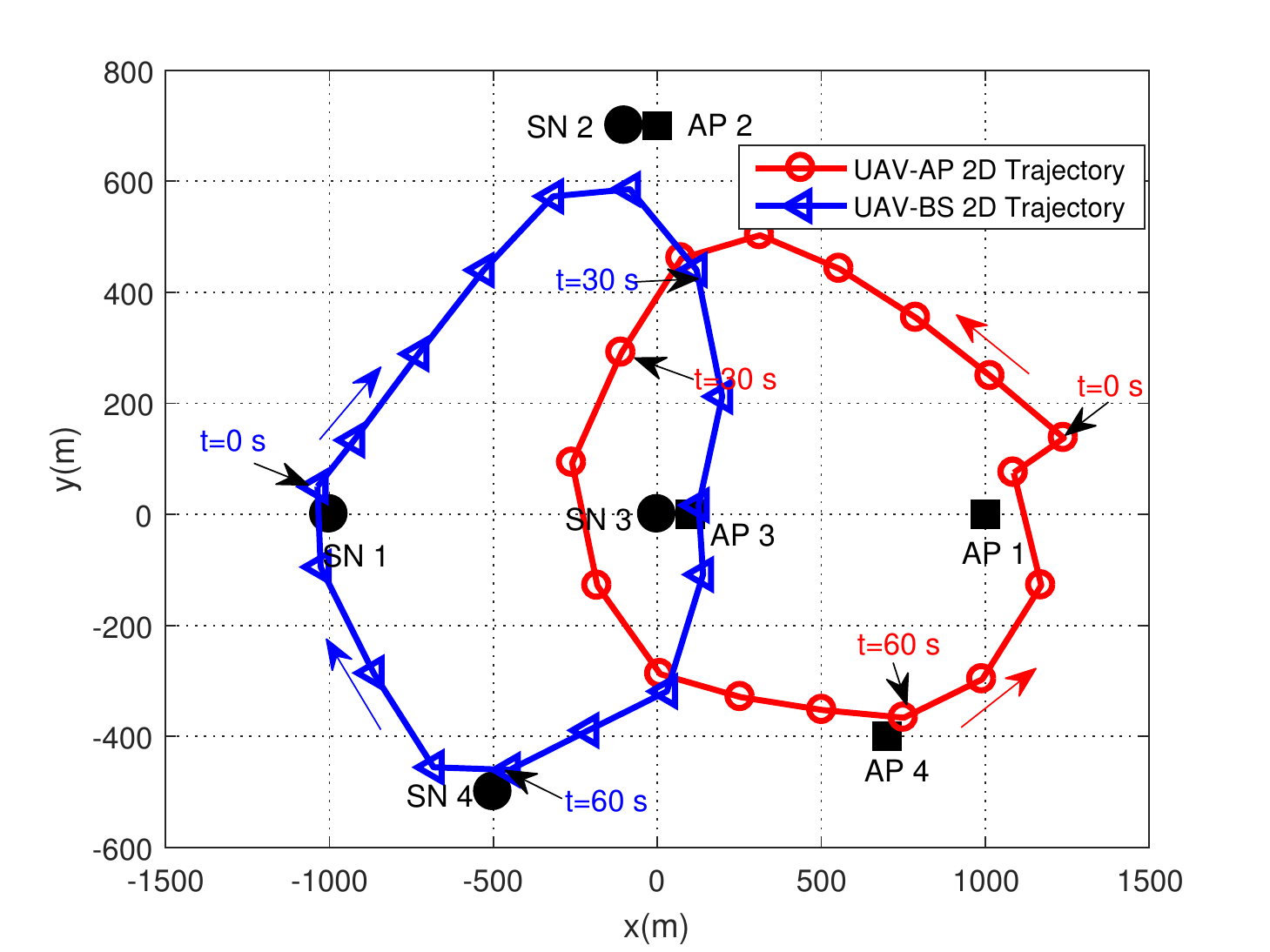}
\end{minipage}
}%
\quad
\centering
\caption{UAV trajectories for  different  weighting  factors  under $T=80s$.  Each trajectory is sampled every 5 seconds with the blue left arrow $\lhd$ marking the UAV-BS trajectory and the red circle $\rm o$ marking the   UAV-AP trajectory.}\label{SMSMfig1}
\end{figure*}

\begin{figure*}[!t]
\centering
\subfigure[$\beta_1=\beta_2=1$.]{
\begin{minipage}[t]{0.45\linewidth}
\centering
\includegraphics[width=2.8in]{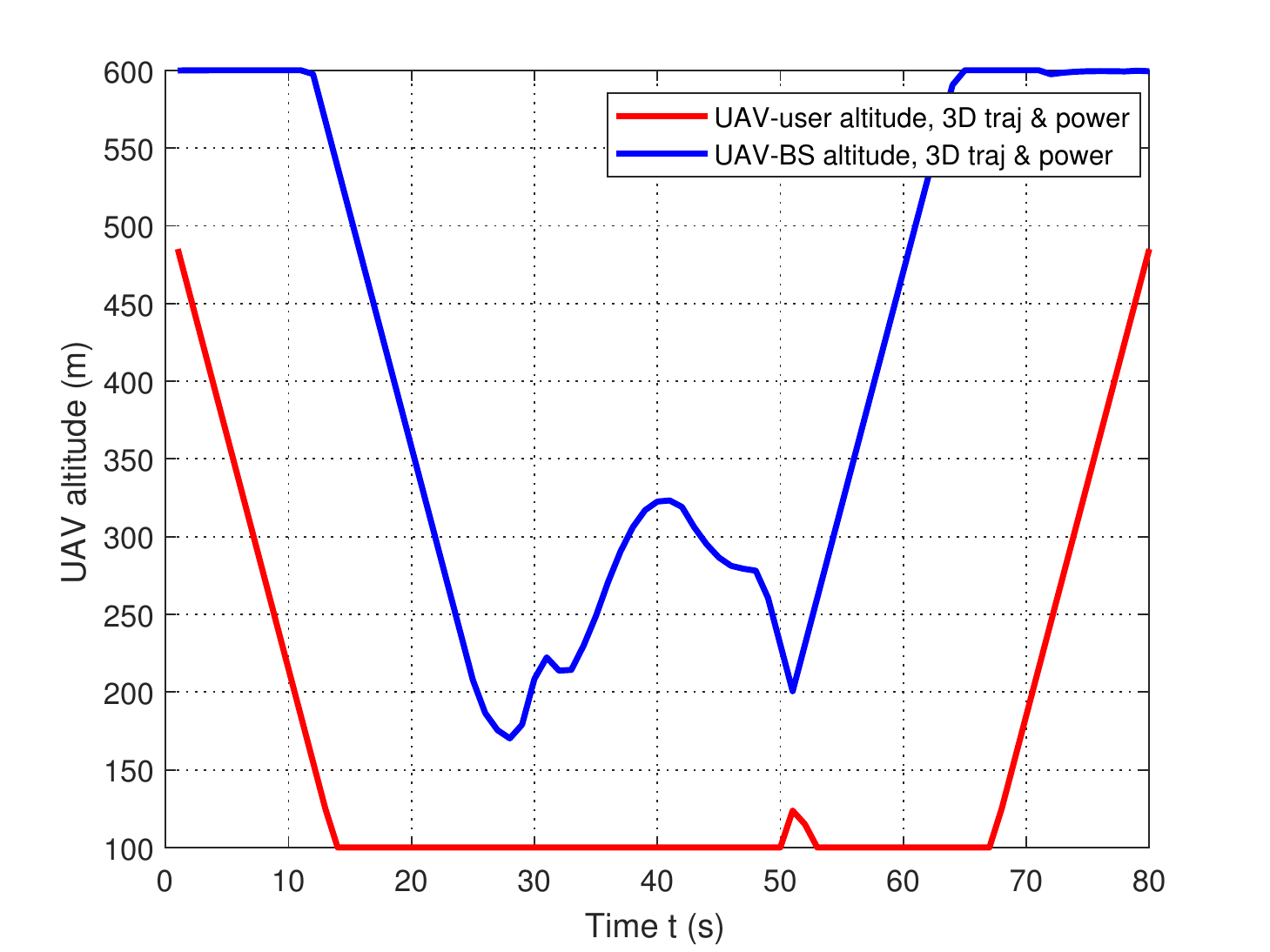}
\end{minipage}%
}%
\quad               
\subfigure[$\beta_1=1, \beta_2=1/10$.]{
\begin{minipage}[t]{0.45\linewidth}
\centering
\includegraphics[width=2.8in]{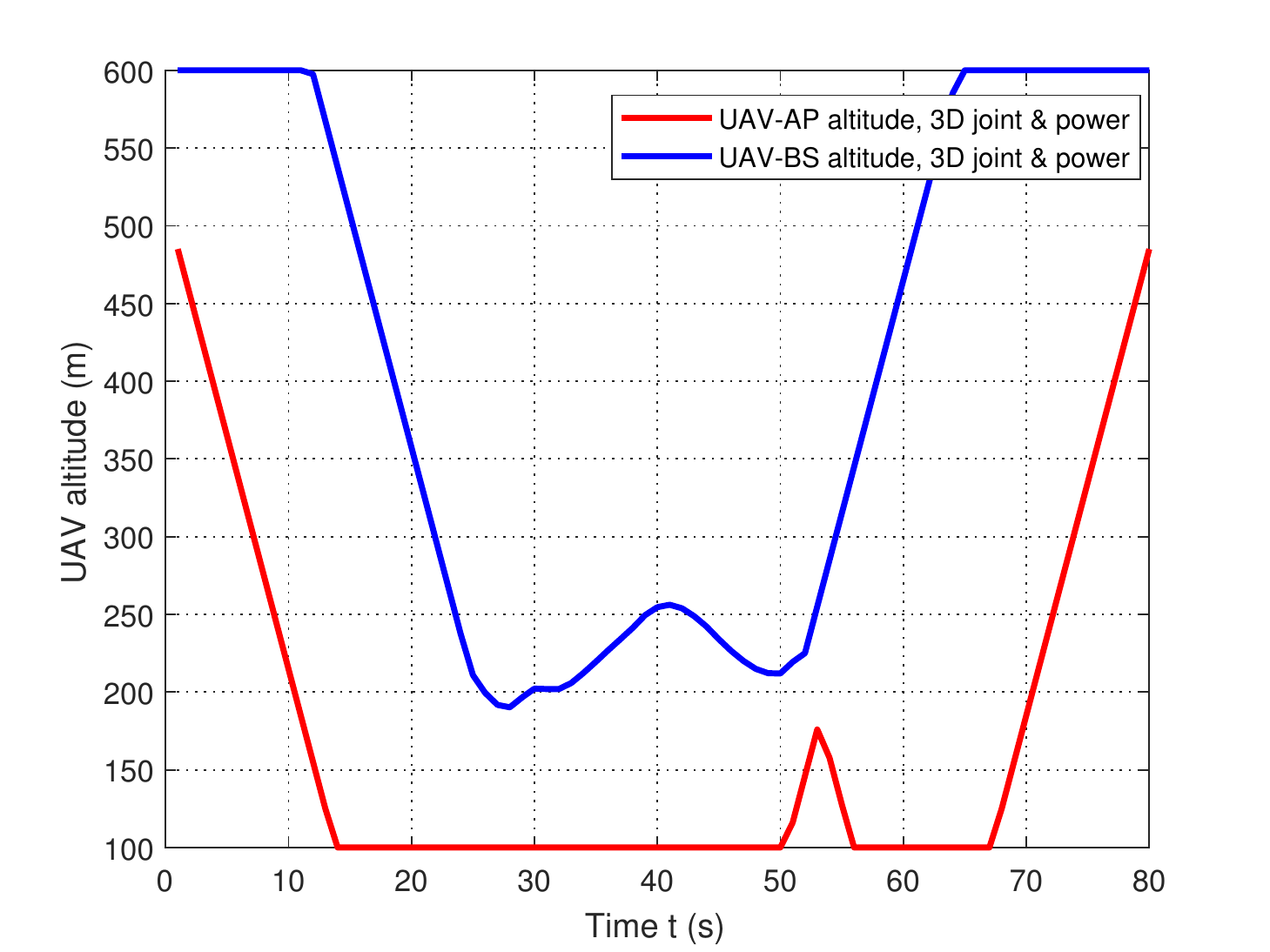}
\end{minipage}
}%
\quad
\centering
\caption{Optimized UAV altitudes for the different weight factors under $T=80s$.}\label{SMSMfig2}
\end{figure*}
In the following,  the 3D UAV trajectory, speed, transmit power, and system throughput  are evaluated under different weighting factors: $\beta_1=\beta_2=1$, and $\beta_1=1, \beta_2=1/10$.
In Fig.~\ref{SMSMfig1}, we show  the optimized UAV trajectories obtained from Algorithm~\ref{alg3} under different weighting factors. It can be observed from Fig.~\ref{SMSMfig1} (a) that both UAVs, i.e., UAV-BS and UAV-AP, sequentially visit SNs and APs, respectively. This is because that the path loss between the UAV and the  ground node would be significantly reduced as the UAV moves closer to the ground node, thereby improving the system throughput.  One can also see  the trajectory that UAV flies from one ground node to another ground node is not a straight line. The reasons have two aspects. One the one hand, the AP not only receives the desired signal from the UAV-AP, but also suffers from interference from the SNs. On the other hand, the UAV-BS not only collects desired data from the SNs, but also encounters interference from the UAV-AP. Therefore, the two UAVs trajectories need to be carefully designed so as to mitigate the strong interference. For $\beta_1=1, \beta_2=1/10$ in Fig.~\ref{SMSMfig1} (b), we can obtain the similar  trajectories  as the case of $\beta_1=\beta_2=1$. However, we can observe from Fig.~\ref{SMSMfig1} (b) that the trajectory that UAV-BS flies from one ground node to another ground node is nearly straight. This is because the UAV-BS networks has a high priority over the UAV-AP networks when $\beta_1>\beta_2$. Therefore, the UAV-BS tends to maximize its own network throughput by optimizing the UAV-BS trajectory. Moreover, the corresponding UAV altitude for the different weighting factors is plotted in Fig.~\ref{SMSMfig2}. It can be observed that both UAVs descend to reduce the path loss, thereby improving the system throughput. This also indicates that the UAV altitude  provides an additional  degree of freedom for performance enhancement.

\begin{figure*}[!t]
\centering
\subfigure[$\beta_1=1, \beta_2=1$.]{
\begin{minipage}[t]{0.45\linewidth}
\centering
\includegraphics[width=2.8in]{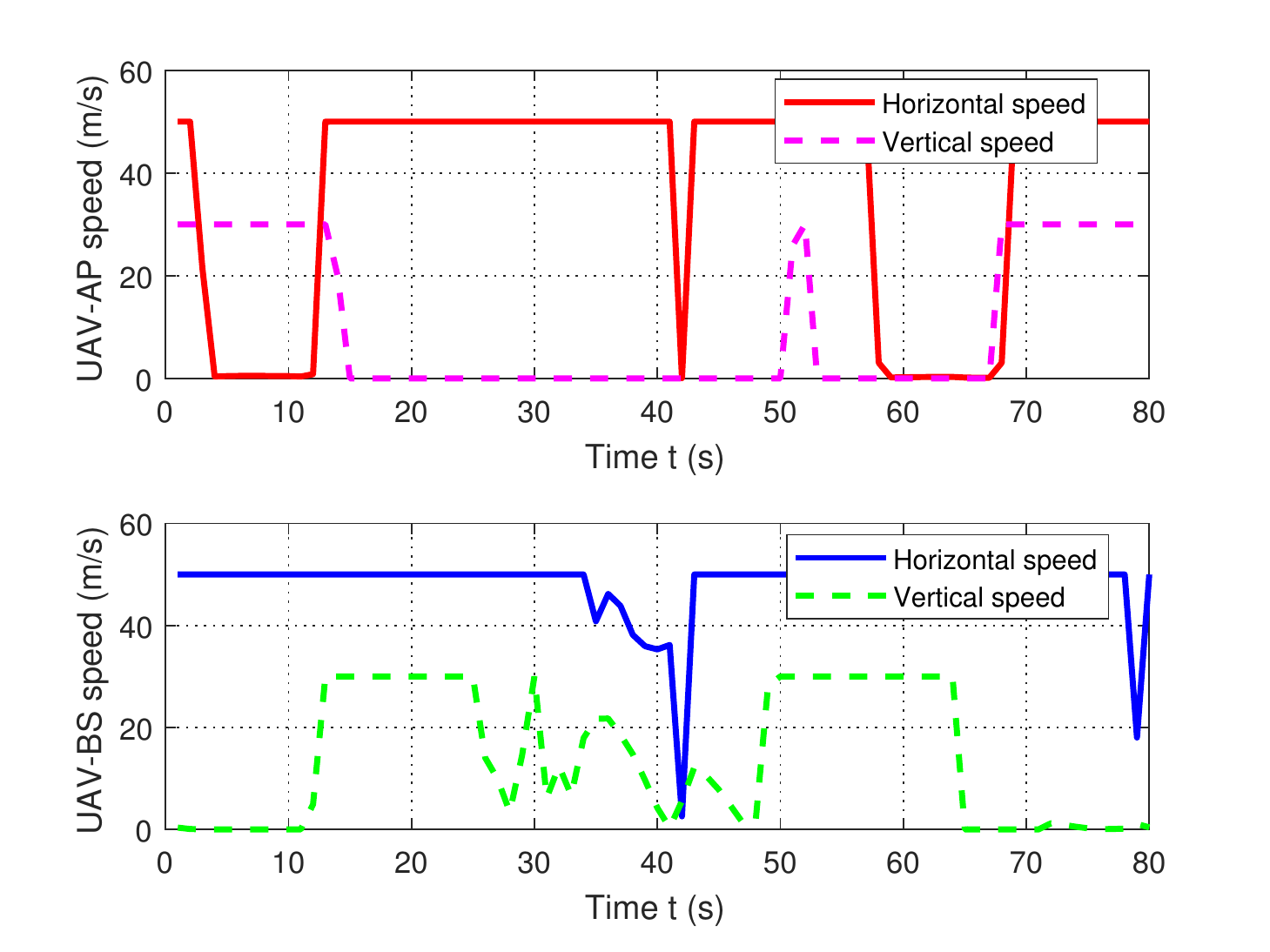}
\end{minipage}%
}%
\quad               
\subfigure[$\beta_1=1, \beta_2=1/10$.]{
\begin{minipage}[t]{0.45\linewidth}
\centering
\includegraphics[width=2.8in]{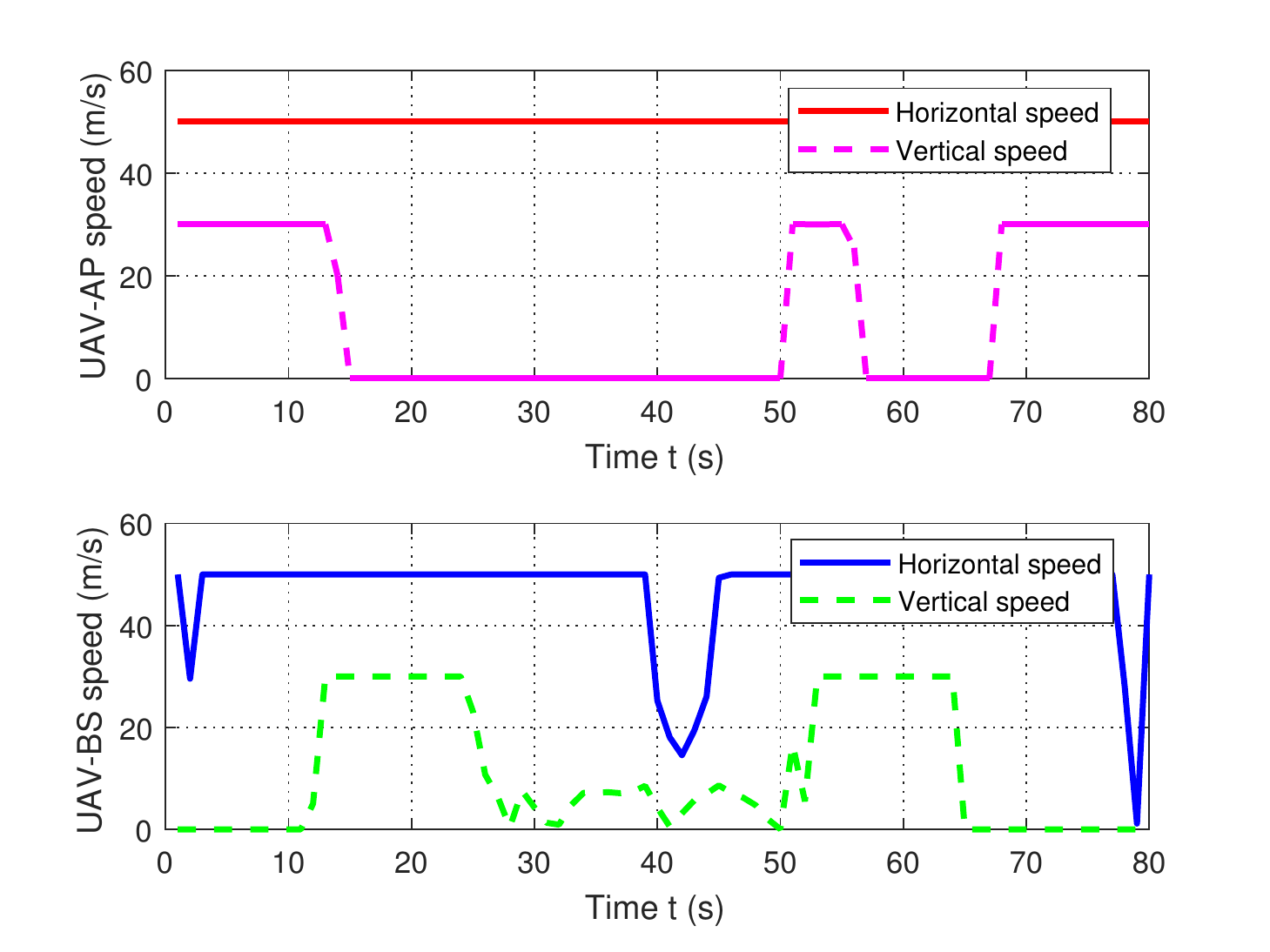}
\end{minipage}
}%
\quad
\centering
\caption{Optimized UAV speed for the different weight factors under $T=80s$.}\label{SMSMfig3}
\end{figure*}

In Fig.~\ref{SMSMfig3}, the  UAV speed for the different weighting factors under $T=80s$ is plotted. It is observed from Fig.~\ref{SMSMfig3} (a) that the UAV flies either with nearly maximum horizontal speed or zero.  This is because exploiting the UAV altitude  provides an additional  degree of freedom for performance enhancement. Unlink Fig.~\ref{SMSMfig3} (a), the UAV-AP flies with the maximum horizontal UAV speed for nearly the  whole period $T$  in Fig.~\ref{SMSMfig3} (b). This is because that the weighting factor $\beta_2=1/10$ for the UAV-AP networks is  small.

\begin{figure*}[htbp]
\centering
\subfigure[$\beta_1=1, \beta_2=1$.]{
\begin{minipage}[t]{0.45\linewidth}
\centering
\includegraphics[width=2.8in]{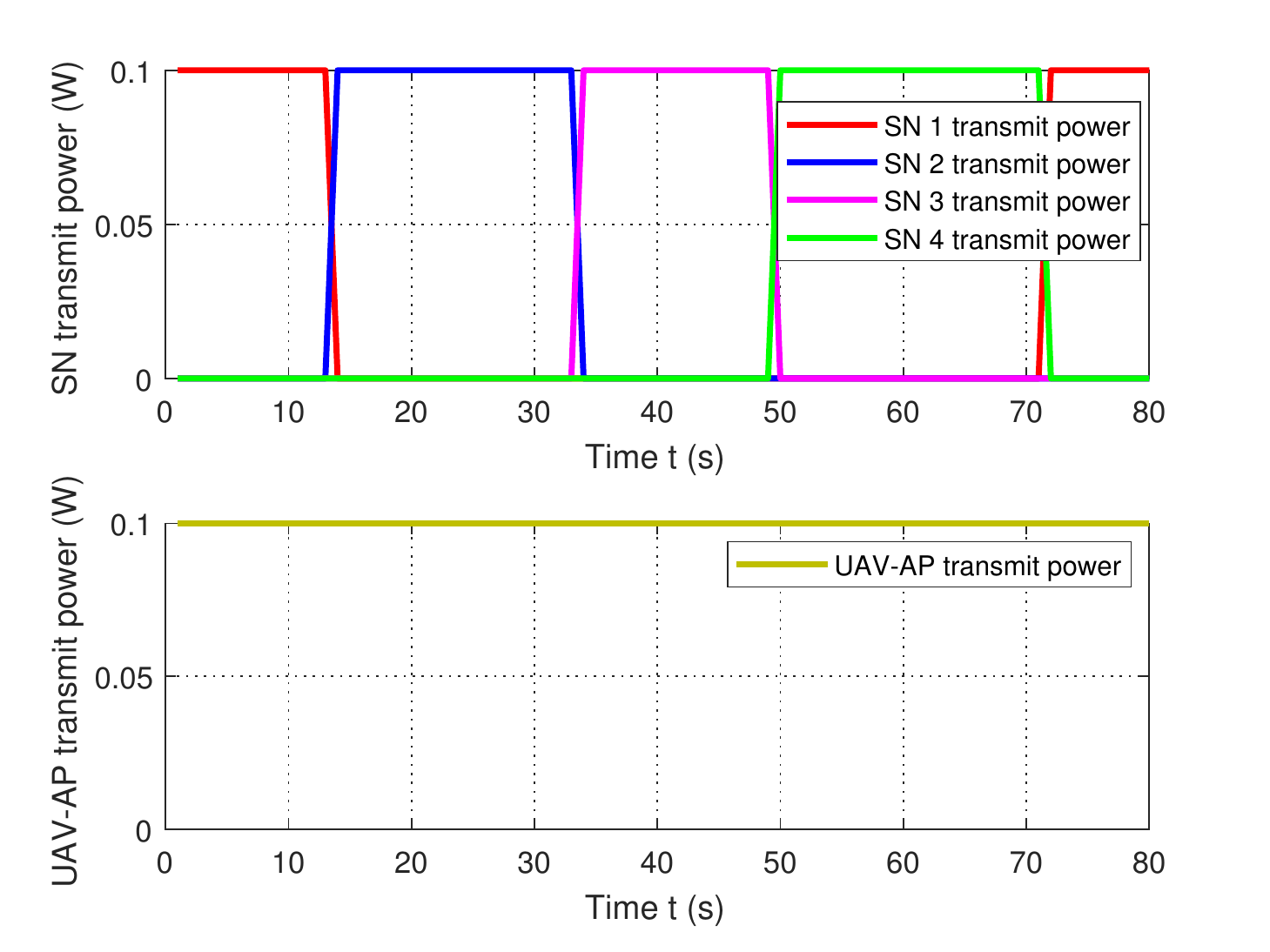}
\end{minipage}%
}%
\subfigure[$\beta_1=1, \beta_2=1/10$.]{
\begin{minipage}[t]{0.45\linewidth}
\centering
\includegraphics[width=2.8in]{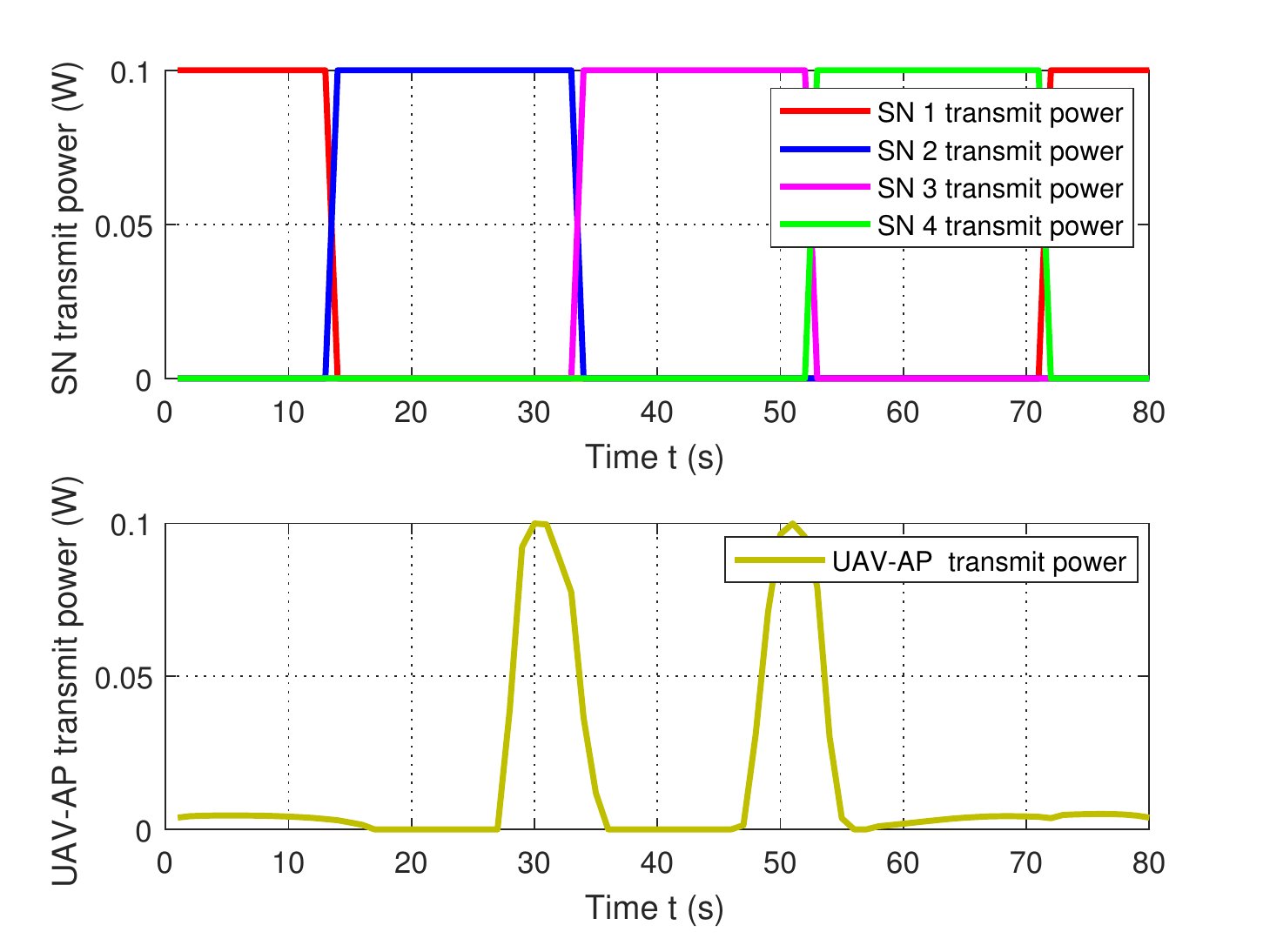}
\end{minipage}%
}%
\quad               
\centering
\caption{Optimized UAV-AP/SN transmit power for the different weight factors under $T=80s$.}\label{SMSMfig4}
\end{figure*}

Fig.~\ref{SMSMfig4} shows  the  UAV-AP/SN transmit power  for the different weighting factors under $T=80s$
It can be seen from Fig.~\ref{SMSMfig4} (a) that the UAV-AP always transmits with maximum power, and the SNs transmit
either with maximum power or zero. However, in Fig.~\ref{SMSMfig4} (b), the UAV-AP transmits with maximum power only from  $t=28s$ to $t=35s$ and $t=48s$ to $t=55s$, and no power is transmitted during other times.  This is expected since with a smaller $\beta_2$, the UAV-AP keeps mute  will  alleviate the interference imposed on the UAV-BS, thereby improving the UAV-BS system throughput.

\begin{figure*}[!t]
\centering
\subfigure[UAV-AP based system throughput]{
\begin{minipage}[t]{0.33\linewidth}
\centering
\includegraphics[width=2.8in]{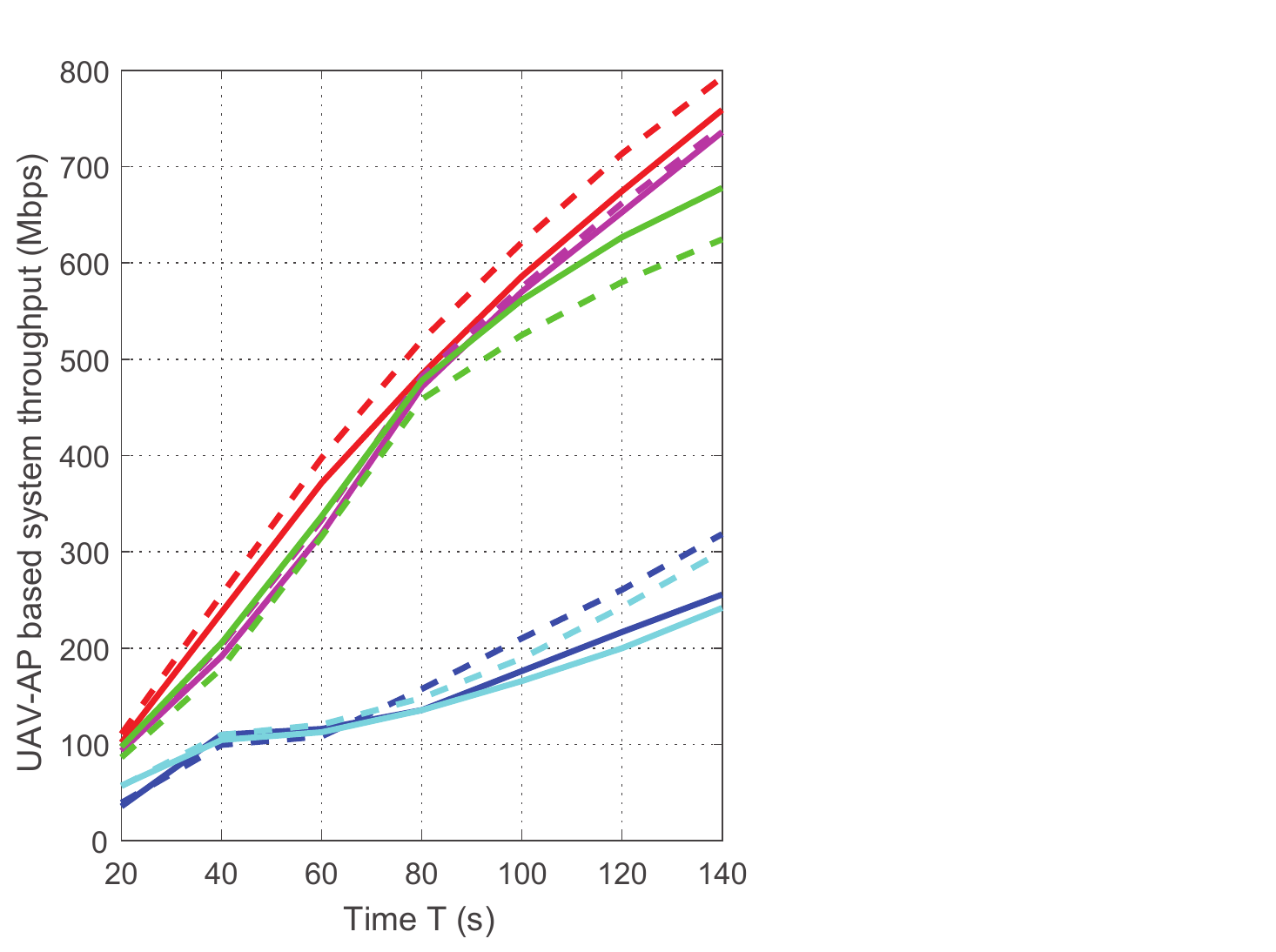}
\end{minipage}%
}%
\subfigure[UAV-BS based system throughput]{
\begin{minipage}[t]{0.33\linewidth}
\centering
\includegraphics[width=2.8in]{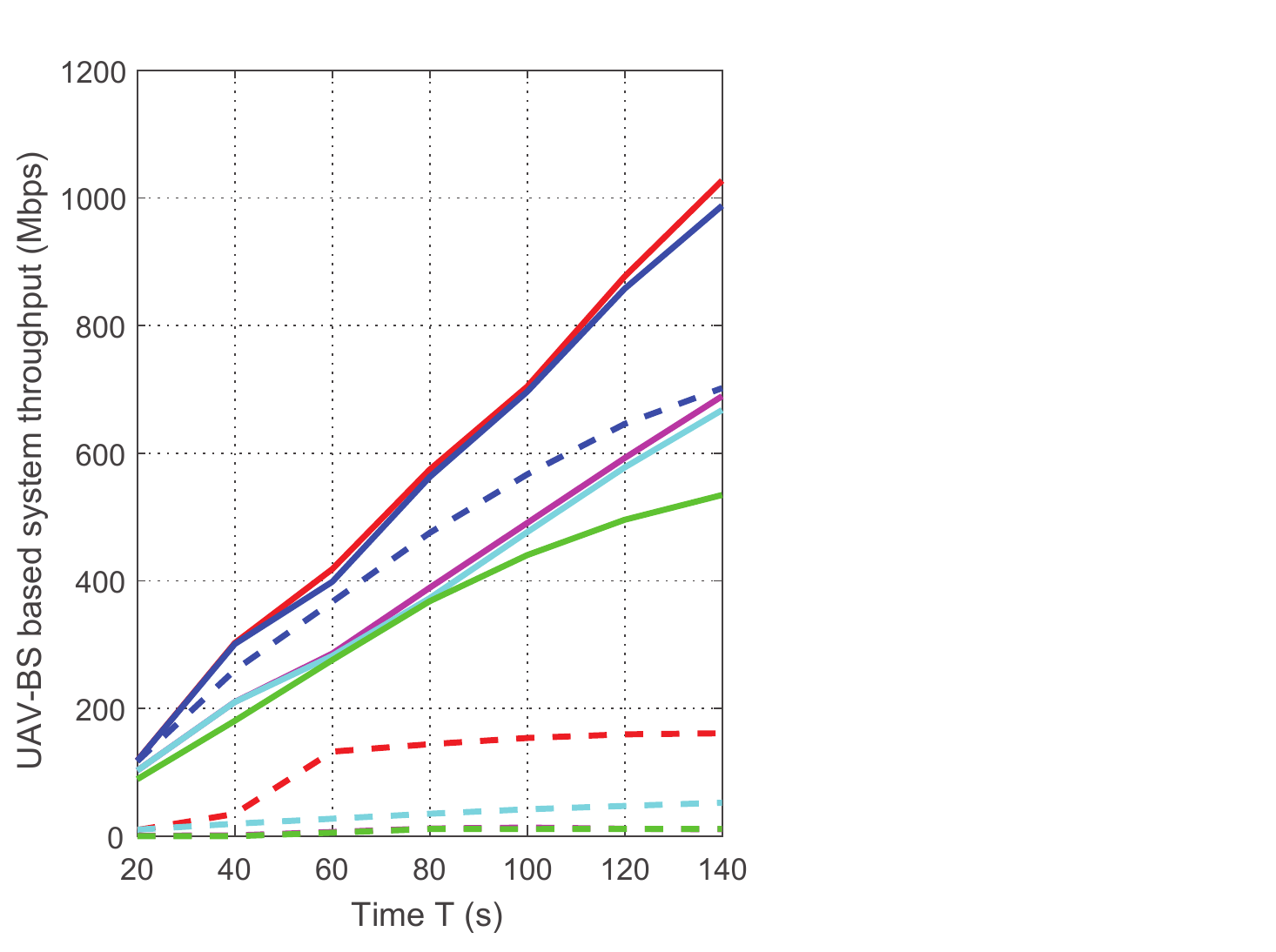}
\end{minipage}%
}
\subfigure[Total system throughput]{
\begin{minipage}[t]{0.33\linewidth}
\centering
\includegraphics[width=2.8in]{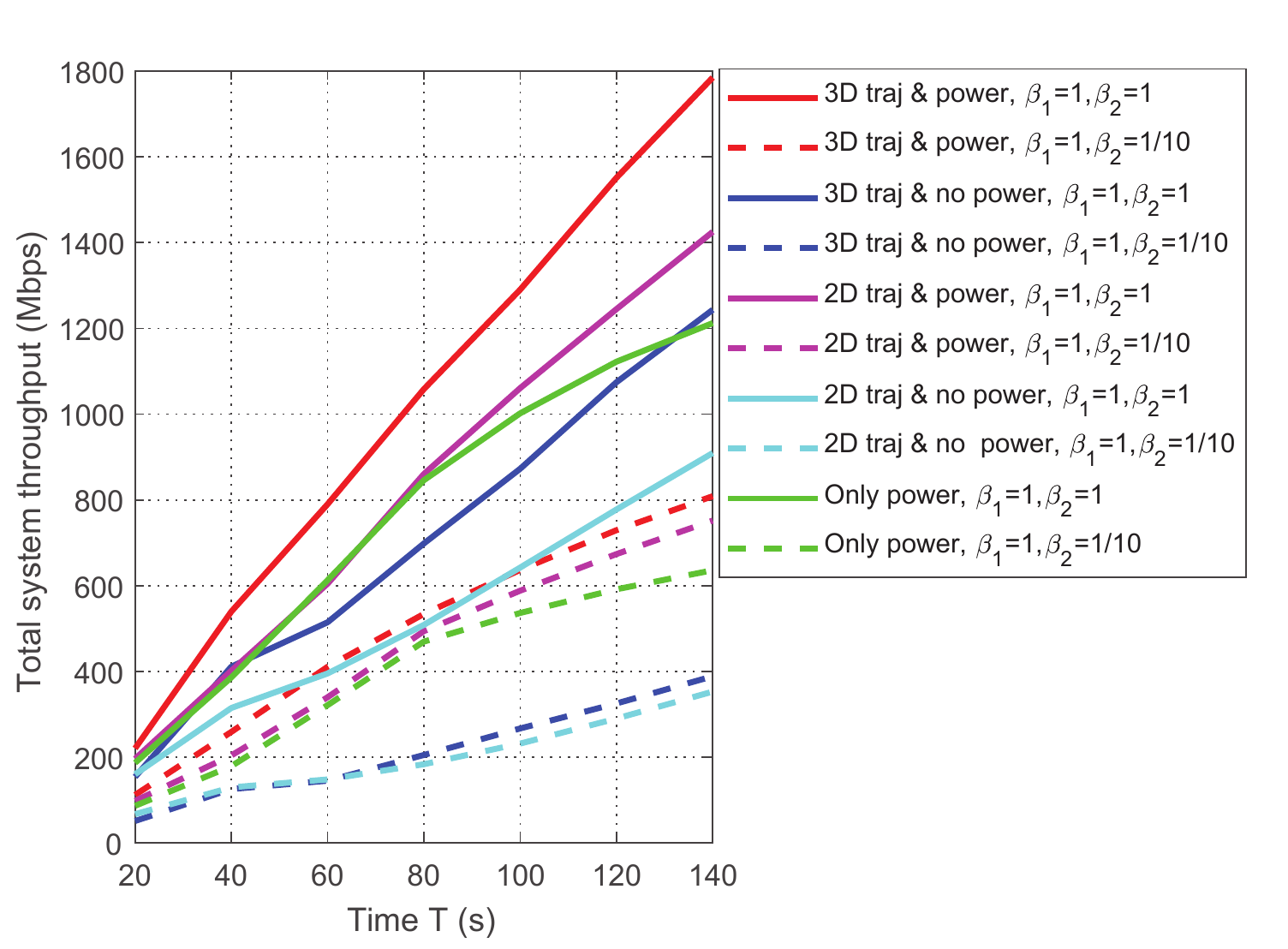}
\end{minipage}
}%
\quad
\centering
\caption{System throughput versus period $T$ for  different  benchmarks under different weighting  factors.}\label{SMSMfig6}
\end{figure*}
In Fig.~\ref{SMSMfig6}, we compare our proposed design with  different benchmarks  for the different weighting  factors in terms of system throughput. The UAV-AP, UAV-BS, and the total system throughput are respectively shown in Fig.~\ref{SMSMfig6} (a), Fig.~\ref{SMSMfig6} (b), and Fig.~\ref{SMSMfig6} (c).  First, we see that our proposed scheme significantly outperforms the other benchmarks as shown in Fig.~\ref{SMSMfig6} (c). For example, for period $T=120s$ and $\beta_2=1$, the total system throughput for  the proposed scheme  is  1551$\rm Mbps$, which is 30\% higher than for ``3D traj \& no power" (1074 $\rm Mbps$), 20\% higher than ``2D traj \& power" (1245 $\rm Mbps$), 50\% higher than ``2D traj \& no power" (777 $\rm Mbps$), and 27\% higher than the ``only power" (1122 $\rm Mbps$) algorithm. This demonstrates the superiority of the proposed scheme. In addition, the benefits of system performance gains can be obtained via UAV altitude optimization, which again confirms that $3\rm D$ trajectory optimization outperforms $2\rm D$ trajectory optimization.  Second, we can observe from Fig.~\ref{SMSMfig6} (b) that ``3D traj \& power" with $\beta_1=\beta_2=1$ achieves a higher system throughput than ``3D traj \& power" with $\beta_1=1, \beta_2=1/10$. This is because  the UAV-BS network has a high priority compared to the UAV-AP network when $\beta_1>\beta_2$, thereby significantly improving the UAV-BS system  throughput.
\section{Conclusion}
This paper studied the UAV-aided simultaneous uplink and downlink transmission networks, where one UAV-AP migrated  data to the APs, and one  UAV-BS collected  data from the SNs. First, we considered a  scenario where  the two UAV trajectories were  pre-determined, and the system throughput was maximized by leveraging the  polyblock outer approximation method. Second, we developed a 3D trajectory and communication design approach for maximizing the system throughput, and a locally optimal solution was achieved by applying the successive convex approximation method. Numerical results  showed  that the proposed successive convex approximation  method   achieved nearly the same system throughput compared with  the polyblock outer approximation method when the UAVs trajectory were pre-determined. In addition,  compared with the benchmarks, a significant system throughput gain was obtained by optimizing the 3D UAV trajectory as well as the transmit power. This work can be extended by considering  multiple UAV-BS and  UAV-AP. The additional  interference caused by additional UAV-BS and  UAV-AP should be carefully managed in order to maximize the system throughput.
\appendices
\section{Proof of Theorem~\ref{theorem1} } \label{appendix1}
Let $f_1\left( {{z}} \right) = {\log _2}\left( {1 + {z}} \right)$ and $f_2\left( {{z}} \right) = {\log _2}\left( {1 + {1 \over {{z}}}} \right)$,  ${z} > 0$. It can be readily checked that $f_1\left( {{z}} \right)$ is concave with respect to $z$ and $f_2\left( {{z}} \right)$ is convex  with respect to $z$, based on the Jensen's inequality \cite{boyd2004convex}, which thus leads to the following inequalities
\begin{align}
{\log _2}\left( {1 + {1 \over {{\mathbb E}\left\{ 1\over z \right\}}}} \right) \le {\mathbb E}\left\{ {{{\log }_2}\left( {1 + z} \right)} \right\} \le {\log _2}\left( {1 + {\mathbb E}\left\{ z \right\}} \right). \label{appendix1_const1}
\end{align}
Define $z = {X \over Y}$ $(X>0, Y>0)$, we have
\begin{align}
{\log _2}\left( {1 + {1 \over {{\mathbb E}\left\{ Y \over X \right\}}}} \right) & \le {\mathbb E}\left\{ {{{\log }_2}\left( {1 + {X \over Y}} \right)} \right\} \notag\\
& \le {\log _2}\left( {1 + {\mathbb E}\left\{ X \over Y \right\}} \right). \label{appendix1_const2}
\end{align}
If $X$ and $Y$ are independent with each other ($X>0$ and $Y>0$), we have
\begin{align}
{\mathbb E}\left\{ {{X \over Y}} \right\} = {\mathbb E}\left\{ X \right\}{\mathbb E}\left\{ {{1 \over Y}} \right\} \ge {{{\mathbb E}\left\{ X \right\}} \over {{\mathbb E}\left\{ Y \right\}}}, \label{appendix1_const3}
\end{align}
where the inequality holds due to the convexity of  function ${1 \over Y}$ for  $Y>0$ and Jensen's inequality.
Based on \eqref{appendix1_const3}, we can derive
\begin{align}
{\log _2}\left( {1 + {1 \over {{\mathbb E}\left\{ {{Y \over X}} \right\}}}} \right)& \le {\log _2}\left( {1 + {{\mathbb E\left\{ X \right\}} \over {\mathbb E\left\{ Y \right\}}}} \right) \notag\\
&\le {\log _2}\left( {1 +\mathbb E\left\{ {{X \over Y}} \right\}} \right).\label{appendix1_const4}
\end{align}
Comparing \eqref{appendix1_const2} and \eqref{appendix1_const4}, we can see that $\mathbb E\left\{ {{{\log }_2}\left( {1 + {X \over Y}} \right)} \right\}$ and ${\log _2}\left( {1 + {{\mathbb E\left\{ X \right\}} \over {\mathbb E\left\{  Y \right\}}}} \right)$ have the same lower bound and upper bound results. In addition, for the special case $X=0, Y>0$, we have $\mathbb E\left\{ {{{\log }_2}\left( {1 + {X \over Y}} \right)} \right\} = {\log _2}\left( {1 + {\textstyle{{\mathbb  E\left\{ X \right\}} \over {\mathbb  E\left\{ Y \right\}}}}} \right)=0$. As a result, we obtain the approximation results in \eqref{theorem1result1}.

\section{Proof of Theorem~\ref{theorem2} } \label{appendix2}
We prove Theorem~\ref{theorem2} in  two steps. In the first step,  we show that the optimal SN transmit power  (UAV-AP transmit power) for  problem \eqref{P3} results in  at most one SN (AP) being active in each time slot. Define ${\check R}_k^s[n]$ as
\begin{align}
{\check R}_k^s [n] ={\log _2}\left( {1 + \frac{{{h_k}\left[ n \right]\tilde p_k^s\left[ n \right]}}{{M\sum\nolimits_{i \ne k}^K {\tilde p_i^s\left[ n \right]}  + \sum\nolimits_{l = 1}^L {f\left[ n \right]\tilde p_l^u\left[ n \right]}  + {\sigma ^2}}}} \right). \label{appendix2_const1}
\end{align}
Suppose that  more than one SN is active,  and   assume that   there is $K_1$ number of  SNs whose transmit power are non-zero, define  $\tilde p_k^s[n] \ne 0$  for $k=1,...,K_1$ $(2 \le K_1\le K)$  and $\tilde p_k^s[n] = 0$  for $k=K_1+1,...,K$. Obviously,  for $\forall k \in \{K_1+1,...,K\}$, ${\check R}_k^s [n] = 0$. For $\forall k \in \{1,...,K_1\}$ with a sufficiently large penalty factor $M \gg 1$,  $M{\sum\limits_{i \ne k}^K {\tilde p_i^s\left[ n \right]} } \to \infty$. Thus, $\sum\limits_{k = 1}^K {{\check R}_k^s} \left[ n \right] = 0$ at any time slot $n$. Suppose that there is  only one SN whose transmit power is  non-zero. We assume that  $\tilde p_1^s[n] \ne 0$   and $\tilde p_k^s[n] = 0$  for $k=2,...,K$. We have
\begin{align}
\sum\limits_{k = 1}^K {R_k^s} \left[ n \right] &= {\log _2}\left( {1 + \frac{{{h_1}\left[ n \right]\tilde p_1^s\left[ n \right]}}{{\sum\nolimits_{l = 1}^L {f\left[ n \right]\tilde p_l^u\left[ n \right]}  + {\sigma ^2}}}} \right)\notag\\
&\overset{a}{=}{\log _2}\left( {1 + {{{h_1}\left[ n \right]\tilde p_1^s\left[ n \right]} \over {f\left[ n \right]\tilde p_l^u\left[ n \right] + {\sigma ^2}}}} \right) > 0, \label{appendix2_const1-2}
\end{align}
where $(a)$ holds since at most one AP is scheduled at any time slot $n$ shown  in the later (we assume that AP $l$ is scheduled here without loss of generality). Therefore, we can declare that at most one  SN is active in order to maximize  \eqref{P3}.
Similarly, define
\begin{align}
{\check R}_l^u [n]={\log _2}\left( {1 + \frac{{{g_l}[n]\tilde p_l^u\left[ n \right]}}{{M\sum\nolimits_{i \ne l}^L {\tilde p_i^u\left[ n \right] + } \sum\nolimits_{k = 1}^K {{{\tilde h}_{k,l}}\tilde p_k^s\left[ n \right] + {\sigma ^2}} }}} \right). \label{appendix2_const2}
\end{align}
It is not difficult to  verify  that at most one  AP can be active in order to maximize \eqref{P3}, based on the same derivation as in \eqref{appendix2_const1}.

In the second step, we show that  \eqref{P3} is equivalent to  \eqref{P2}. First, it can be easily seen that  the optimal solution to problem \eqref{P2} is feasible for  problem \eqref{P3}   with the same objective value. Second, based on  the first step, we  see that the optimal solution to problem \eqref{P3} is also  feasible for problem \eqref{P2}  with same objective value.
This thus completes the proof of  Theorem~\ref{theorem2}.
\bibliographystyle{IEEEtran}
\bibliography{TCOM3Dtrajectory}
\begin{IEEEbiography}[{\includegraphics[width=1in,height=1.25in,clip,keepaspectratio]{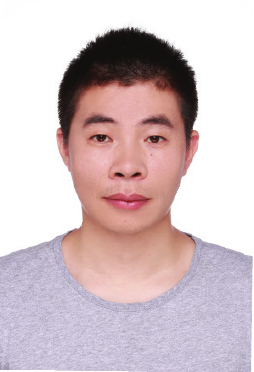}}]{Meng Hua}
	received the  M.S. degree in electrical and information engineering from Nanjing University of Science and Technology, Nanjing, China, in 2016. Since September 2016, he is currently working towards the Ph.D. degree in School of Information Science and Engineering, Southeast University, Nanjing, China. His current research interests include UAV assisted communication, intelligent reflecting surface (IRS), backscatter communication,  energy-efficient wireless communication, X-connectivity, cognitive radio network, secure transmission, and optimization theory.
\end{IEEEbiography}

\begin{IEEEbiography}[{\includegraphics[width=1in,height=1.25in,clip,keepaspectratio]{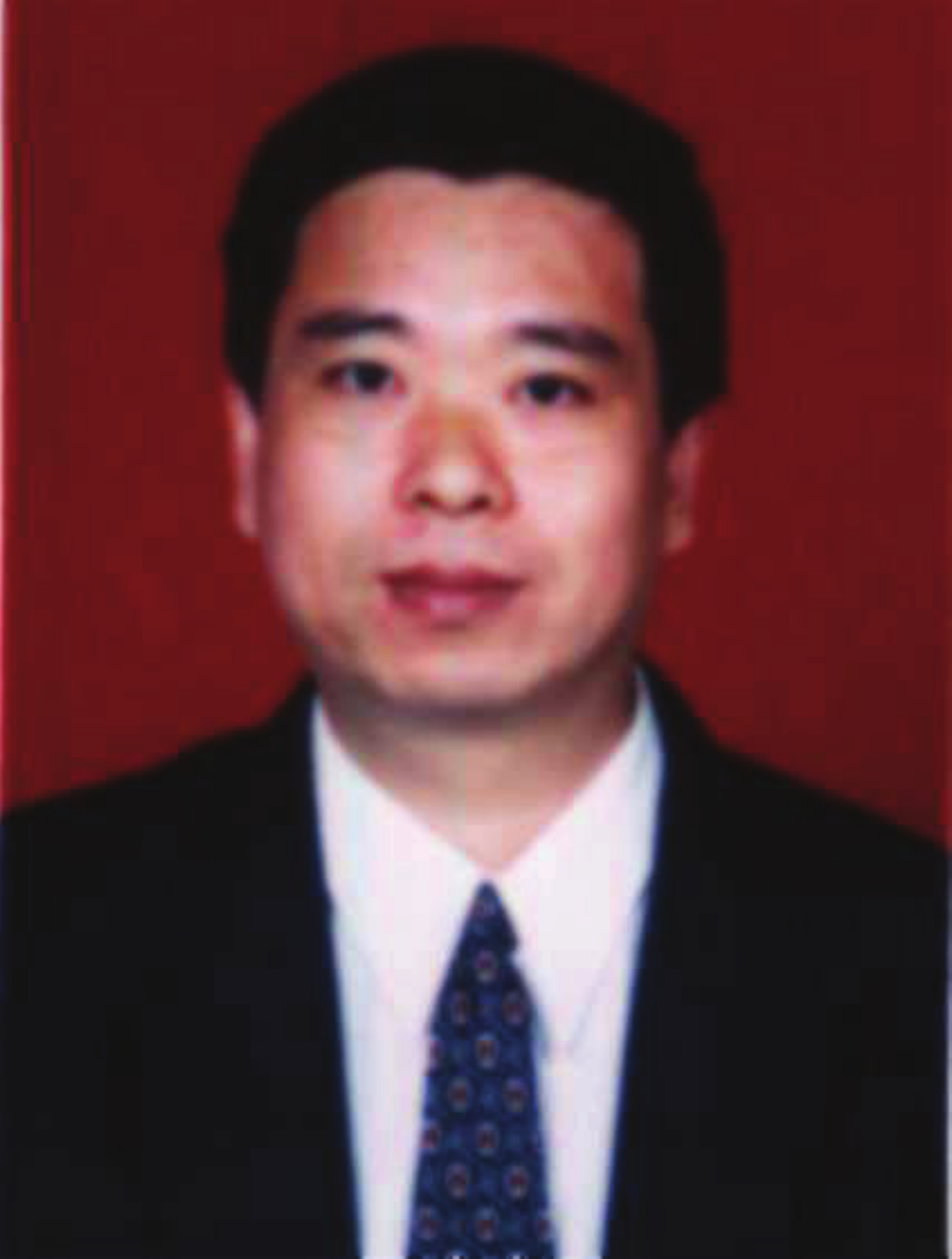}}]{Luxi Yang}
	(M'96-SM'17) received the M.S. and Ph.D. degrees in electrical engineering from Southeast University, Nanjing, China, in 1990 and 1993, respectively. Since 1993, he has been with the Department of Radio Engineering, Southeast University, where he is currently a Full Professor of information systems and communications, and the Director of the Digital Signal Processing Division. He has authored or co-authored of two published books and more than 200 journal papers, and holds 50 patents. His current research interests include signal processing for wireless communications, MIMO communications, intelligent wireless communications, and statistical signal processing. He received the first and second class prizes of science and technology progress awards of the State Education Ministry of China in 1998, 2002, and 2014. He is currently a member of Signal Processing Committee of the Chinese Institute of Electronics.
\end{IEEEbiography}

\begin{IEEEbiography}[{\includegraphics[width=1in,height=1.25in,clip,keepaspectratio]{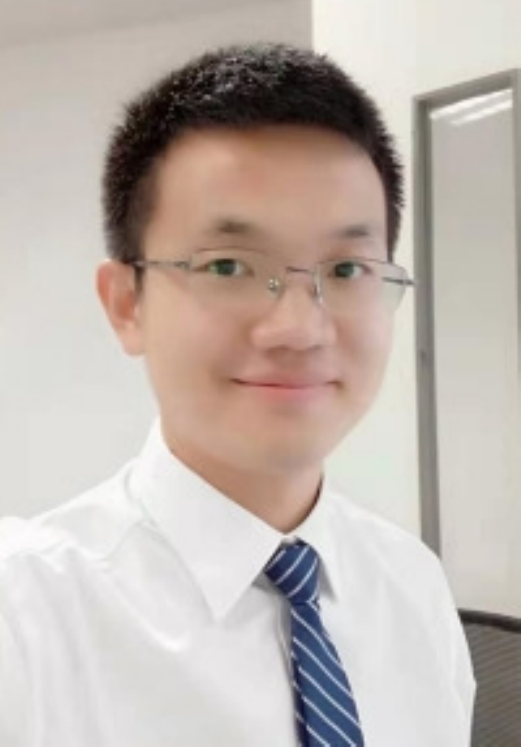}}] {Qingqing Wu} (S'13-M'16) received the B.Eng. and the Ph.D. degrees in Electronic Engineering from South China University of Technology and Shanghai Jiao Tong University (SJTU) in 2012 and 2016, respectively. He is currently an Assistant Professor in the Department of Electrical and Computer Engineering at the University of Macau, China, and also with the State key laboratory of Internet of Things for Smart City. He was a Research Fellow in the Department of Electrical and Computer Engineering at National University of Singapore. His current research interest includes intelligent reflecting surface (IRS), unmanned aerial vehicle (UAV) communications, and MIMO transceiver design. He has published over 60 IEEE journal and conference papers.
	
He was the recipient of the IEEE WCSP Best Paper Award in 2015, the Outstanding Ph.D. Thesis Funding in SJTU in 2016, the Outstanding Ph.D. Thesis Award of China Institute of Communications in 2017. He was the Exemplary Editor of IEEE Communications Letters in 2019 and the Exemplary Reviewer of several IEEE journals. He serves as an Associate Editor for IEEE Communications Letters and IEEE Open Journal of Communications Society. He is the Lead Guest Editor for IEEE Journal on Selected Areas in Communications on ``UAV Communications in 5G and Beyond Networks", and the Guest Editor for IEEE Open Journal on Vehicular Technology on ``6G Intelligent Communications" and IEEE Open Journal of Communications Society on ``Reconfigurable Intelligent Surface-Based  Communications for 6G Wireless Networks". He is the workshop co-chair for ICC 2019 and ICC 2020 workshop on ``Integrating UAVs into 5G and Beyond", and the workshop co-chair for GLOBECOM 2020 workshop on ``Reconfigurable Intelligent Surfaces for Wireless Communication for Beyond 5G". 
\end{IEEEbiography}

\begin{IEEEbiography}[{\includegraphics[width=1in,height=1.25in,clip,keepaspectratio]{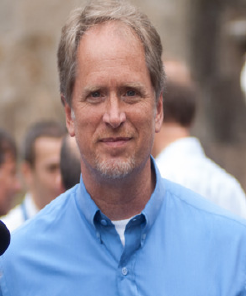}}]{A. Lee Swindlehurst}
	(F'04) received the B.S. (1985) and M.S. (1986) degrees in Electrical Engineering from Brigham Young University (BYU), and the PhD (1991) degree in Electrical Engineering from Stanford University. He was with the Department of Electrical and Computer Engineering at BYU from 1990-2007, where he served as Department Chair from 2003-06.  During 1996-97, he held a joint appointment as a visiting scholar at Uppsala University and the Royal Institute of Technology in Sweden. From 2006-07, he was on leave working as Vice President of Research for ArrayComm LLC in San Jose, California. Since 2007 he has been a Professor in the Electrical Engineering and Computer Science Department at the University of California Irvine, where he served as Associate Dean for Research and Graduate Studies in the Samueli School of Engineering from 2013-16. During 2014-17 he was also a Hans Fischer Senior Fellow in the Institute for Advanced Studies at the Technical University of Munich. In 2016, he was elected as a Foreign Member of the Royal Swedish Academy of Engineering Sciences (IVA). His research focuses on array signal processing for radar, wireless communications, and biomedical applications, and he has over 300 publications in these areas. Dr. Swindlehurst is a Fellow of the IEEE and was the inaugural Editor-in-Chief of the IEEE Journal of Selected Topics in Signal Processing. He received the 2000 IEEE W. R. G. Baker Prize Paper Award, the 2006 IEEE Communications Society Stephen O. Rice Prize in the Field of Communication Theory, the 2006 and 2010 IEEE Signal Processing Society's Best Paper Awards, and the 2017 IEEE Signal Processing Society Donald G. Fink Overview Paper Award.
\end{IEEEbiography}

\end{document}